\documentclass[12pt]{article}
\pdfoutput=1

\usepackage{putex}
\usepackage{graphicx}
\usepackage{caption}
\usepackage{amsmath}
\usepackage{amssymb}
\usepackage{array}
\usepackage{multirow}
\usepackage{mathtools}
\usepackage{comment}
\usepackage{subcaption}
\usepackage{epstopdf}
\usepackage{enumerate}
\usepackage{cite}
\usepackage{youngtab}
\usepackage{tensor}
\usepackage{slashed}
\usepackage[aligntableaux=center]{ytableau}
\usepackage[utf8]{inputenc}
\usepackage{rotating}
\usepackage{bigfoot}
\usepackage[
      colorlinks=true,
      linkcolor=blue,
      urlcolor=blue,
      filecolor=black,
      citecolor=red,
      linktocpage=true
      ]{hyperref}

\newcommand{\abs}[1]{\left\lvert #1 \right\rvert}

\newcommand {\be} {\begin {equation}}
\newcommand {\ee} {\end {equation}}

\newcommand {\bes} {\begin {equation*}}
\newcommand {\ees} {\end {equation*}}

\newcommand{\es}[2]{%
  \begin{equation}
    \begin{aligned}
      #2
    \end{aligned}
    \phantomsection\label{#1}%
  \end{equation}%
}

\newcommand{\cA}{{\mathcal A}}

\newcommand{\cF}{{\mathcal F}}
\newcommand{\cG}{{\mathcal G}}

\newcommand{\cN}{{\mathcal N}}

\newcommand   \zb  {\bar{z}}

\newcommand{\bea}{\begin{equation}\begin{aligned}}
\newcommand{\eea}[1]{\label{#1}\end{aligned}\end{equation}}

\newcommand{\beq}{\begin{equation}}
\newcommand{\eeq}{\end{equation}}

\def\ie{\begin{equation}\begin{aligned}}
\def\fe{\end{aligned}\end{equation}}

\numberwithin{equation}{section}

\def\<{\langle}
\def\>{\rangle}

\begin{document}

\preprint{}

\institution{oxford}{${}^{\mu_1}$ Mathematical Institute, University of Oxford,
Woodstock Road, Oxford, OX2 6GG, UK}
\institution{HU}{${}^{\mu_2}$ Jefferson Physical Laboratory, Harvard University, Cambridge, MA 02138, USA}
\institution{HUU}{${}^{\mu_3}$ Center of Mathematical Sciences and Applications, Harvard University, Cambridge, MA 02138, USA}
\institution{stony}{${}^{\mu_4}$ Simons Center for Geometry and Physics, SUNY, Stony Brook, NY 11794, USA}

\title{Gluon scattering in AdS at finite string coupling from localization}

\authors{Connor Behan,${}^{\mu_1}$\worksat{\oxford}  Shai M.~Chester,${}^{\mu_2, \mu_3}$\worksat{\HU}\worksat{\HUU} and Pietro Ferrero${}^{\mu_4}$\worksat{\stony}  }

\abstract{
We consider gluons scattering in Type IIB string theory on AdS$_5\times S^5/\mathbb{Z}_2$ in the presence of D7 branes, which is dual to the flavor multiplet correlator in a certain 4d $\mathcal{N}=2$ $USp(2N)$ gauge theory with $SO(8)$ flavor symmetry and complexified coupling $\tau$. We compute this holographic correlator in the large $N$ and finite $\tau$ expansion using constraints from derivatives of the mass deformed sphere free energy, which we compute to all orders in $1/N$ and finite $\tau$ using supersymmetric localization. In particular, we fix the $F^4$ higher derivative correction to gluon scattering on AdS at finite string coupling $\tau_s=\tau$
in terms of Jacobi theta functions, which feature the expected relations between the $SL(2,\mathbb{Z})$ duality and the $SO(8)$ triality of the CFT, and match it to the known flat space term.
We also use the flat space limit to compute $D^2F^4$ corrections of the correlator at finite $\tau$ in terms of a non-holomorphic Eisenstein series.
At weak string coupling, we find that the AdS correlator takes a form which is remarkably similar to that of the flat space Veneziano amplitude.
}
\date{}

\maketitle

\tableofcontents

\section{Introduction}
\label{intro}

The AdS/CFT correspondence \cite{Witten:1998qj,Gubser:1998bc,Maldacena:1997re} relates quantum gravity in AdS$_{d+1}$ to a conformal field theory (CFT) in $d$ dimensions. In the paradigmatic example \cite{Maldacena:1997re}, Type IIB string theory on AdS$_5\times S^5$ is dual to $SU(N)$ $\mathcal{N}=4$ super-Yang-Mills (SYM), where $N$ is inversely related to the string length $\ell_s$ and the complexified string coupling $\tau_s$ is dual to the complexifed gauge coupling $\tau$. This duality is believed to be exact, so we should be able to study non-perturbative string theory from CFT. In practice, however, it has been difficult to study holographic CFTs like $\mathcal{N}=4$ SYM at finite coupling, and thus difficult to probe string theory at finite $\tau_s$ from AdS/CFT.

Progress on AdS/CFT at finite string coupling has emerged in recent years by combining two non-perturbative CFT methods: the conformal bootstrap \cite{Rattazzi:2008pe} and supersymmetric localization \cite{Pestun:2007rz}. Crossing symmetry can be applied to correlation functions of holographic CFTs such as $\mathcal{N}=4$ SYM to fix their functional dependence in the large $N$ limit in terms of just a few parameters at each order \cite{Rastelli:2017udc,Alday:2014tsa}. Some of these parameters can then be computed as a function of CFT parameters such as $\tau$ using supersymmetric localization \cite{Binder:2019jwn}. We can then take the flat space limit \cite{Penedones:2010ue} of the AdS amplitude dual to the CFT correlator, and compare to the S-matrix in the corresponding quantum gravity theory.

So far, this program has been applied to graviton scattering, which is dual to the stress tensor correlator, and for string theory duals is related to closed string scattering. For instance, certain integrals of the 4d $\mathcal{N}=4$ stress tensor correlator were related to derivatives of the mass deformed sphere free energy $F(m)$ \cite{Binder:2019jwn,Chester:2020dja}, which can be expressed as an $N$-dimensional matrix model using supersymmetric localization \cite{Pestun:2007rz}. The matrix model expressions were computed to all orders in $1/N$ and finite $\tau$, and then used to fix the coefficients of the large $N$ correlator to several sub-leading orders in terms of functions of $\tau$ that were invariant under the $SL(2,\mathbb{Z})$ duality group \cite{Chester:2019jas,Chester:2020vyz}. In the flat space limit, these finite $\tau$ coefficients were found to precisely match protected higher derivative corrections to the Type IIB S-matrix in the small $\ell_s$ expansion at finite $\tau_s$, which had been computed before using $SL(2,\mathbb{Z})$ and perturbative string calculations \cite{Green:1997as,Green:1998by,Green:1997tv,Alday:2021vfb}. Similar results have also been found for 3d CFTs dual to Type IIA/M-theory \cite{Chester:2018aca,Binder:2018yvd,Binder:2019mpb,Alday:2021ymb,Alday:2022rly}, and 6d CFTs\footnote{In this case, instead of localization one can use non-trivial constraints from the 2d chiral algebra \cite{Beem:2014kka}.} dual to M-theory \cite{Chester:2018dga,Alday:2020tgi}.

In this paper, we generalize this program to gluon scattering on branes, which is dual to the flavor multiplet correlator, and is related to open string scattering. In particular, we consider Type IIB string theory with four D7 branes, an O7 plane, and $N$ D3 branes, which can be engineered from F-theory on a $D_4$ singularity \cite{Sen:1996vd,Banks:1996nj}. This is the simplest F-theory construction, as $\tau_s$ can take any value, is independent of the torus, and has a weak coupling limit. The near horizon limit of the D3 branes is the orientifold AdS$_5\times S^5/\mathbb{Z}_2$. The dual CFT is a 4d $\mathcal{N}=2$ $USp(2N)$ gauge theory with four fundamental hypermultiplets, an antisymmetric hypermultiplet, and an $SO(8)$ flavor symmetry. This theory is a conformal manifold with one complex parameter $\tau$, which is related to the UV complexified gauge coupling $\tau_\text{UV}$ as \cite{Hollands:2010xa,Douglas:1996js}
\es{UVtoIR}{
e^{4\pi i \tau_\text{UV}}=16\frac{\theta_2(\tau/2)^4}{\theta_3(\tau/2)^4}\,,\qquad \tau_\text{UV} = \frac{\theta}{2\pi}+i\frac{4\pi}{g_\text{YM}^2}\,,
}
where $\theta_i$ are Jacobi theta functions (defined in \eqref{jacobi}), and $\tau$ transforms naturally under the $SL(2,\mathbb{Z})$ duality group of the conformal manifold. The CFT parameters are related to bulk parameters as
\es{adscft}{
\frac{L^4}{\ell_s^4}=8\pi g_s N\,,\qquad \tau_s\equiv\frac{\chi}{2\pi}+\frac{i}{g_s}=\tau\,,
}
where $L$ is the AdS radius. We consider the flavor multiplet four-point function, which is dual to gluons scattering on the D7 branes. At large $N$ (\textit{i.e.} small $\ell_s$), the gluons scatter on AdS$_5\times S^3$, which is the fixed point locus of the D7 branes on the orientifold. We will compute this holographic correlator at large $N$ and finite $\tau=\tau_s$ in a few steps.

First we will review the constraints of the analytic bootstrap on the large $N$ correlator \cite{Zhou:2018ofp,Alday:2021odx,Alday:2021ajh}. The leading tree gluon and graviton exchange terms are totally fixed. The next corrections in $1/N$ come from the higher derivative contact term $F^4$, the 1-loop gluon exchange term, and a $\log N$ term that regulates the logarithmic divergence of the 1-loop term, all of which are fixed up to three coefficients. The last term we will consider is the further subleading $D^2F^4$ contact term, which is fixed up to five coefficients. 

Next, we will constrain the correlator using localization. In \cite{Chester:2022sqb}, integrals of the flavor multiplet correlator were related to derivatives of the mass deformed sphere free energy $F(\mu_i)$ for any 4d $\mathcal{N}=2$ CFT. For our case of $SO(8)$ flavor symmetry, there are three such independent constraints, which are related by triality of $SO(8)$. Following \cite{Beccaria:2021ism,Beccaria:2022kxy}, we compute the matrix model expression for $F(\mu_i)$ to all orders in $1/N$ and finite $\tau$. We find that all $1/N$ corrections are independent of $\tau$ except the leading term, which is written in terms of Jacobi theta functions of $\tau$. For instance, one such mass derivative is
\es{mIntro}{
-\partial_{\mu_1}^2\partial_{\mu_2}^2 F\big|_{\mu=0}=-2\log[\tau_2 |\theta_3(\tau)\theta_4(\tau)|^2]+h(N)+O(e^{-N})\,,
}
where $\tau\equiv\tau_1+i\tau_2$, $h(N)$ is a known $\tau$-independent function, and the other mass derivatives are related by the interplay between $SO(8)$ triality and the $SL(2,\mathbb{Z})$ duality group \cite{Sen:1996vd,Seiberg:1994aj}. We can use these constraints to fix three coefficients at each order in $1/N$, which fixes the $F^4$ coefficients in terms of these modular functions, and the 1-loop and $\log N$ contact terms in terms of $\tau$-independent numbers.

We then take the flat space limit of the holographic correlator and compare to gluons scattering on D7 branes. At weak string coupling, this open string scattering is given by the Veneziano amplitude \cite{Polchinski:1998rr} 
\es{V}{
\mathcal{A}_V=&-(2\pi)^5{\ell_s^4}\left[\frac{V(s,t,\ell_s)}{s\,t}\text{tr}(T^AT^BT^CT^D)+\frac{V(t,u,\ell_s)}{t\,u}\text{tr}(T^AT^DT^BT^C)\right.\\
&\left.\qquad\qquad\quad+\frac{V(s,u,\ell_s)}{s\,u}\text{tr}(T^AT^CT^DT^B)\right]\,,\quad V(s,t,\ell_s)\equiv\frac{\Gamma[1-\ell_s^2s]\Gamma[1-\ell_s^2t]}{\Gamma[1-\ell_s^2(s+t)]}\,,\\
}
where $s,t,u=-s-t$ are Mandelstam variables and $T^A$ are generators of the $SO(8)$ bulk gauge symmetry. The Veneziano amplitude is expanded at low energy as 
\es{Vsmall}{
V(s,t,\ell_s)=1-\zeta(2)\ell_s^4st+\zeta(3)\ell_s^6stu+O(\ell_s^8)\,,
}
where the genus-zero $F^4$ term at order $\ell_s^4$ matches the flat limit of the holographic correlator. The next correction is the genus-zero $D^2F^4$ term at order $\ell_s^6$, which we compare to the flat limit of the holographic correlator and combine with the localization constraints to fix all five coefficients of the AdS $D^2F^4$ term. The genus-zero (\textit{i.e.} leading $1/N$) holographic correlator $M$ then takes a very similar form as the flat space Veneziano amplitude:\footnote{For other recent work related to the AdS Veneziano amplitude, see \cite{Glew:2023wik}. We thank the authors for coordinating the submission with us.}
\es{AdSV}{
{M}=-\frac{2}{N}&\left[\frac{\tilde{V}(s,t,\lambda)}{(s-2)\,(t-2)}\text{tr}(T^AT^BT^CT^D)+\frac{\tilde{V}(t,{u},\lambda)}{({t}-2)\,(u-2)}\text{tr}(T^AT^DT^BT^C)\right.\\
&\left.+\frac{\tilde{V}(s,{u},\lambda)}{(s-2)\,({u}-2)}\text{tr}(T^AT^CT^DT^B)\right]+O(1/N^2)\,,\qquad \frac{1}{\lambda}\equiv \frac{1}{g_\text{YM}^2N}+\frac{\log2}{2\pi^2N}\,,
}
where $s,t,u=6-s-t$ here are AdS Mellin amplitude variables, $T^A$ are now generators of the $SO(8)$ flavor symmetry, the IR $\lambda$ is shifted relative to the UV as in \eqref{UVtoIR}, and $\tilde{V}(s,t)$ to the order we computed is
\es{AdSV2}{
\tilde{V}(s,t)=1-\frac{24\zeta(2)}{\lambda}(s-2)(t-2)+\frac{192\zeta(3)}{\lambda^{3/2}}(s-2)(t-2)({u}-2)+O(\lambda^{-2})\,.
}
This closely resembles the small $\ell_s$ expansion of the flat space term $V(s,t)$ in \eqref{Vsmall} except with $s,t\to s-2,t-2$.

The flat space $F^4$ correction was also computed at finite string coupling $\tau_s$ using duality to heterotic string theory in \cite{Bachas:1997mc,Gutperle:1999xu,Bachas:1997xn,Foerger:1998kw,Bianchi:1998vq,Gava:1999ky,Kiritsis:2000zi,Lerche:1998gz,Lerche:1998nx}, as well as by computing D(-1) instanton amplitudes in Type IIB string theory in \cite{Gava:1999ky,Billo:2009gc,Billo:2010mg}.\footnote{More precisely, only the $\tau_s$-dependent terms were computed by comparison to the heterotic string, since the heterotic string is dual to Type IIB where the transverse space is a torus, while in our case the transverse space is flat. Since $F^4$ is protected, the $\tau_s$ dependence is expected to be the same on either space, but the $\tau_s$-independent terms may differ, so we only compare $\tau_s$-dependent terms in our holographic correlator. We thank Ofer Aharony for discussion on this.} We find that it precisely matches the flat limit of the AdS $F^4$ term that we computed from localization, which is a check of AdS/CFT at finite string coupling. The $D^2F^4$ correction is also protected, and we show following \cite{Wang:2015jna,Lin:2015ixa} how $SL(2,\mathbb{Z})$ invariance and weak coupling results fix it in flat space at finite $\tau_s$ in terms of a rank $3/2$ non-holomorphic Eisenstein series $E_{3/2}(\tau_s)$. We then combine the constraint from flat space with the localization constraints to fix the AdS $D^2F^4$ coefficient also in terms of $E_{3/2}(\tau_s)$.

The rest of this paper is organized as follows. In Section \ref{gscat} we discuss kinematic constraints from superconformal symmetry on the flavor multiplet correlator, as well as relations between the integrated correlator and $F(\mu_i)$. In Section \ref{pertString} we use these integrated constraints as well as the flat limit to constrain the large $N$ expansion of the holographic correlator in the strong 't Hooft coupling expansion, which is dual to weak string coupling. In Section \ref{NPString}, we similarly constrain the correlator in the large $N$ and finite $\tau$ limit, which is dual to finite string coupling. We conclude in Section \ref{conc} with a review of our results and a discussion of future directions. Technical details of the calculations are given in the various Appendices.

 \section{Gluon scattering in AdS$_5$}
\label{gscat}

The main object of study in this work is the moment map four-point function in a 4d $\mathcal{N}=2$ CFT with $SO(8)$ symmetry, which is dual to gluon scattering in the F-theory construction. We begin by reviewing general constraints from the $\mathcal{N}=2$ superconformal algebra following \cite{Beem:2014zpa}. We then discuss integrated constraints in both position and Mellin space from the mass deformed sphere free energy \cite{Chester:2022sqb}. Everything discussed in this section is completely non-perturbative.

\subsection{Setup}
\label{setup}

The 4d $\cN=2$ CFT we consider has R-symmetry $SU(2)_R\times U(1)_R$ and flavor symmetries $SU(2)_L$ and $SO(8)$. We consider the flavor multiplet, whose superprimary is the moment map operator $\phi^A(y,x)$, which is a Lorentz scalar of dimension $\Delta=2$ that is a singlet under $SU(2)_L$ and transforms in the adjoint $\bf28$ of $SO(8)$ with index $A$ and the adjoint of $SU(2)_R$ with spinor polarizations $y$. The conformal and global symmetries restrict the 4-point of this operator to be
\es{phiExp1}{
\langle \phi^A(y_1,x_1) \phi^B(y_2,x_2) \phi^C(y_3,x_3) \phi^D(y_4,x_4) \rangle = \frac{\langle y_1,y_2\rangle^2 \langle y_3,y_4\rangle^2}{x_{12}^4x_{34}^4}G^{ABCD}(U,V;w)\,,
}
where we define the cross ratios
\es{w}{
  U \equiv \frac{{x}_{12}^2 {x}_{34}^2}{{x}_{13}^2 {x}_{24}^2} \,, \qquad
   V \equiv \frac{{x}_{14}^2 {x}_{23}^2}{{x}_{13}^2 {x}_{24}^2} \,, \qquad w=\frac{\langle y_1,y_2\rangle \langle y_3,y_4\rangle}{\langle y_1,y_3\rangle \langle y_2,y_4\rangle}\,, 
}
with $x_{12}\equiv x_1-x_2$ and $\langle y_1,y_2\rangle=y^\alpha_1y^\beta_2\varepsilon_{\alpha\beta}$ for $\alpha,\beta=1,2$. We can furthermore impose the flavor symmetry by expanding in projectors $P_r^{ABCD}$ for each flavor irrep $r$ as
\es{flavorProj}{
G^{ABCD}(U,V;w)=\sum_{r\in{\bf 28}\otimes{\bf28}}G_r(U,V;w)P_r^{ABCD}\,,
}
for the flavor irreps
\es{tensso8}{
{\bf28}\otimes{\bf28}={\bf1}\oplus {\bf28}\oplus {\bf35_v}\oplus {\bf35_c}\oplus {\bf35_s}\oplus {\bf300}\oplus {\bf350}\,,
}
where $SO(8)$ triality permutes the three 35-dimensional irreps. We use the same basis of $SO(8)$ tensor structure as in \cite{Chester:2022sqb}:
\begin{align}\label{tbas}
\begin{split}
\mathtt{t}_i=&\left(\delta^{AB}\delta^{CD},\quad \delta^{AC}\delta^{BD},\quad \delta^{AD}\delta^{BC},\quad 
f^{ACE}f^{BDE},\quad f^{ADE}f^{BCE},\right.\\
&\left. \text{tr}(T^AT^BT^CT^D),\quad \frac{1}{4\cdot 4!}\epsilon_{a_1a_2b_1b_2c_1c_2d_1d_2} T^A_{a_1a_2}T^B_{b_1b_2}T^C_{c_1c_2}T^D_{d_1d_2}\right)\,,
\end{split}
\end{align}
where $i=1, \dots, 7$ runs over the seven independent tensor structures, $a=1, \dots, 8$ are fundamental indices, the generators in the fundamental representation are normalized as
\begin{align}
\text{tr}(T^AT^B)=2\delta^{AB}\,,
\end{align}
and the structure constants as
\begin{align}
[T^A,T^B]=i\,f^{ABC}\,T^C\,.
\end{align}
The projectors in this basis are
\begin{align}\label{Pbas}
\begin{split}
P_{\mathbf{1}} & =\left(\begin{array}{lllllll}
\frac{1}{28} & 0 & 0 & 0 & 0 & 0 & 0
\end{array}\right) \,\cdot \mathtt{t}\\
P_{\mathbf{2 8}} & =\left(\begin{array}{lllllll}
0 & 0 & 0 & -\frac{1}{12} & \frac{1}{12} & 0 & 0
\end{array}\right) \cdot \mathtt{t}\\
P_{\mathbf{3 5}_ \mathbf{c}} & =\left(\begin{array}{lllllll}
0 & \frac{1}{12} & \frac{1}{12} & 0 & \frac{1}{12} & -\frac{1}{12} & \frac{1}{2}
\end{array}\right) \cdot \mathtt{t}\\
P_{\mathbf{3 5}_{\mathbf{s}}} & =\left(\begin{array}{lllllll}
0 & \frac{1}{12} & \frac{1}{12} & 0 & \frac{1}{12} & -\frac{1}{12} & -\frac{1}{2}
\end{array}\right)\cdot \mathtt{t} \\
P_{\mathbf{3 5}_ \mathbf{v}} & =\left(\begin{array}{lllllll}
-\frac{1}{12} & 0 & 0 & \frac{1}{12} & -\frac{1}{12} & \frac{1}{6} & 0
\end{array}\right)\cdot \mathtt{t} \\
P_{\mathbf{3 5 0}} & =\left(\begin{array}{lllllll}
0 & -\frac{1}{2} & \frac{1}{2} & \frac{1}{12} & -\frac{1}{12} & 0 & 0
\end{array}\right) \cdot \mathtt{t}\\
P_{\mathbf{3 0 0}} & =\left(\begin{array}{lllllll}
\frac{1}{21} & \frac{1}{3} & \frac{1}{3} & -\frac{1}{12} & -\frac{1}{12} & 0 & 0
\end{array}\right)\cdot \mathtt{t}\,,
\end{split}
\end{align}
and are normalized as 
\es{projNorm}{
P_r^{ABBA}=\text{dim}(r)\,.
}

The last kinematic constraint comes from the superconformal Ward identity, which we can formally solve by writing $G^{ABCD}(U,V;w)$ as \cite{Dolan:2001tt}
\es{ward4dN2}{
G_{r}(U,V;w)=\frac{z(w-\bar z)f_{r}(\bar z)-\bar z(w-z)f_{r}(z)}{w(z-\bar z)}+\left ( 1-\frac{z}{w} \right ) \left ( 1-\frac{\bar z}{w} \right )\mathcal{G}_{r}(U,V)\,,
}
where $U=z\bar z\,, V=(1-z)(1-\bar z)$ and the reduced correlator $\mathcal{G}$ as well as the holomorphic correlator $f(z)$ are now R-symmetry singlets. The holomorphic correlator $f(z)$ is protected and takes the universal form 
\es{freeGen}{
f^{ABCD}(z)=\delta^{AB}\delta^{CD}+z^2\delta^{AC}\delta^{BD}+\frac{z^2}{(1-z)^2}\delta^{AD}\delta^{CB}+\frac{2z}{k}f^{ACE}f^{BDE}+\frac{2z}{k(z-1)}f^{ADE}f^{BCE}\,,
}
where the flavor central charge $k$ is defined in terms of the canonically normalized current 2-point of the flavor current $J_\mu^A$ as 
\es{kDef}{
\left < J_\mu^A(x) J_\nu^B(0) \right >=k\frac{3\delta^{AB}}{4\pi^4}\frac{I_{\mu\nu}}{x^6}\,,\qquad I_{\mu\nu} \equiv \delta_{\mu\nu} - 2\frac{x_\mu x_\nu}{x^2}\,.
}
When discussing gravitational corrections to gluon scattering, it will be useful to refer to the conformal anomaly $c$ which determines the canonically normalized stress tensor as
\es{cDef}{
\left < T_{\mu\nu}(x) T_{\rho\sigma}(0) \right >=\frac{20 c}{\pi^4 x^8} \left ( I_{\mu\sigma} I_{\nu\rho} + I_{\mu\rho} I_{\nu\sigma} - \frac{1}{2} \delta_{\mu\nu} \delta_{\rho\sigma} \right )\,.
}
For the $USp(2N)$ gauge theory we consider these central charges are \cite{Aharony:2007dj}
\es{kcSO8}{
k=4N\,,\qquad c=\frac{N^2}{2}+\frac{3N}{4}-\frac{1}{12}\,,
}
which for $N=1$ reduce to the $SU(2)\cong USp(2)$ SQCD parameters of \cite{Chester:2022sqb}. 

\subsection{Integrated constraints}
\label{intCon}

We can compute exact constraints on the integrated moment map correlator by considering the mass deformed sphere free energy $F(\mu_i)$, where $i=1,\dots,4$ correspond to the four Cartans of $SO(8)$, which can be computed exactly using supersymmetric localization \cite{Pestun:2016zxk}. As shown in \cite{Chester:2022sqb}, these constraints relate the three independent combinations of mass derivatives to the connected correlator $\mathcal{G}^\text{con}\equiv\mathcal{G}-(\mathtt{t}_1+\mathtt{t}_2 U+\mathtt{t}_3U/V^2)$ as\footnote{In \cite{Chester:2022sqb}, the integrated constraints were written in terms of the interacting part of $\mathcal{G}$, and the integral of the free part was added separately, which was a convenient way of taking into account the holomorphic terms in the full correlator in \eqref{ward4dN2}. For the large $N$ analysis we consider below, the holomorphic terms are already included in the full correlator tree gluon exchange term \cite{Alday:2021odx}, which anyway has a flavor structure that does not contribute to the integrated constraints, so we simply write the integrated correlator in terms of the connected part of $\mathcal{G}$.}
\es{intNew}{
-\partial_{\mu_1}^4 F\big|_{\mu=0}&=k^2I[\cG^\text{con}_1+\cG^\text{con}_2+\cG^\text{con}_3+2\cG^\text{con}_6]\,,\\
  -\partial_{\mu_1}^2\partial_{\mu_2}^2 F\big|_{\mu=0}&=\frac{k^2}{3}I[\cG^\text{con}_1+\cG^\text{con}_2+\cG^\text{con}_3]\,,\\
-\partial_{\mu_1}\partial_{\mu_2}\partial_{\mu_3}\partial_{\mu_4} F \big|_{\mu=0}&=\frac{k^2}{6}I[\cG^\text{con}_7]\,,\\
}
where the integral is defined as
 \es{d4FCombinedAgain}{
  I[\mathcal{G}]\equiv\frac{1}{ \pi}  \int dR\, d\theta\, R^3 \sin^2 \theta
     \frac{\bar D_{1,1,1,1}(U,V)\cG(U, V)}{U^2} 
\bigg|_{\substack{U = 1 + R^2 - 2 R \cos \theta \\
    V = R^2}}\,,
 }
 and here we used the $\mathtt{t}_i$ basis \eqref{tbas} for $\cG^\text{con}_i$.  The definition of $\bar{D}_{1,1,1,1}(U,V)$ is
\es{db1111}{
\bar{D}_{1,1,1,1}(U,V) = \frac{1}{z - \zb} \left ( \log(z\zb) \log \frac{1 - z}{1 - \zb} + 2\text{Li}(z) - 2\text{Li}(\zb) \right )\,.
}
 We sometimes find it convenient to write these constraints in the irrep basis as 
 \es{intOld}{
 \cF_{\bf v}&=k^2I\Big[\frac{1}{21}\cG^\text{con}_{\bf 1} -\frac19\cG^\text{con}_{\bf 35_v} +\frac29\cG^\text{con}_{\bf 35_c} +\frac29\cG^\text{con}_{\bf 35_s} +\frac{20}{21}\cG^\text{con}_{\bf 300}  \Big]\,,\\
  \cF_{\bf c}&=k^2I\Big[\frac{1}{21}\cG^\text{con}_{\bf 1} +\frac29\cG^\text{con}_{\bf 35_v} -\frac19\cG^\text{con}_{\bf 35_c} +\frac29\cG^\text{con}_{\bf 35_s} +\frac{20}{21}\cG^\text{con}_{\bf 300}  \Big]\,,\\
   \cF_{\bf s}&=k^2I\Big[\frac{1}{21}\cG^\text{con}_{\bf 1} +\frac29\cG^\text{con}_{\bf 35_v} +\frac29\cG^\text{con}_{\bf 35_c} -\frac19\cG^\text{con}_{\bf 35_s} +\frac{20}{21}\cG^\text{con}_{\bf 300}  \Big]\,,\\
 }
where we define the mass combinations
\es{Fcsv}{
\mathcal{F}_{\bf v}&\equiv   -4\partial_{\mu_1}^2\partial_{\mu_2}^2 F\big|_{\mu=0}\,,\\
 \mathcal{F}_{\bf c}&\equiv    -\partial_{\mu_1}^4 F\big|_{\mu=0}- \partial_{\mu_1}^2\partial_{\mu_2}^2 F\big|_{\mu=0}+2  \partial_{\mu_1}\partial_{\mu_2}\partial_{\mu_3}\partial_{\mu_4} F \big|_{\mu=0}\,,\\
  \mathcal{F}_{\bf s}&\equiv    -\partial_{\mu_1}^4 F\big|_{\mu=0}- \partial_{\mu_1}^2\partial_{\mu_2}^2 F\big|_{\mu=0}-2  \partial_{\mu_1}\partial_{\mu_2}\partial_{\mu_3}\partial_{\mu_4} F \big|_{\mu=0}\,,\\
}
such that the integrated constraints are naturally permuted by $SO(8)$ triality. In the following sections we will write the correlator in Mellin space as \cite{Alday:2021odx}
\es{mellin}{
\mathcal{G}_\text{con}^{ABCD}(U,V)=\int \frac{ds dt}{(4\pi i)^2}U^{s/2}V^{t/2-2} M^{ABCD}(s,t) \Gamma[2-s/2]^2\Gamma[2-t/2]^2\Gamma[2-{u}/2]^2\,,
}
where $s+t+{u}=6$, and the two integration contours include all poles in $s,t$ but not $u$. If desired, one could also obtain the full Mellin amplitude (the one for the connected part of $G(U, V; w)$) from the reduced Mellin amplitude (what we are using) via a difference equation given in \cite{Alday:2021odx}.\footnote{This is consistent because the first term in \eqref{ward4dN2} is a rational function of the cross ratios and therefore does not have its own Mellin amplitude. The same can be said of free correlators but we have checked that the unique $M^{ABCD}(s, t)$ computes $\mathcal{G}_\text{con}^{ABCD}(U, V)$ rather than its interacting part. Note that in $\mathcal{N} = 4$ SYM, it would be possible to work with the latter by choosing a slightly different contour as done in \cite{Rastelli:2017udc}.}  It was found in \cite{future} that the integral \eqref{d4FCombinedAgain} acts on an arbitrary component $M_r(s,t)$ as
\es{Imel}{
  I[M_r]\equiv -\int \frac{ds dt}{(4\pi i)^2}& \Bigg[M_r(s,t) \Gamma[2-s/2]\Gamma[s/2]\Gamma[2-t/2]\Gamma[t/2]\Gamma[2-{u}/2]\Gamma[{u}/2] \\
 & \times \Big(\frac{H_{\frac s2-1}+H_{1-\frac s2}}{(t-2)(u-2)}+\frac{H_{\frac t2-1}+H_{1-\frac t2}}{(s-2)(u-2)}+\frac{H_{\frac u2-1}+H_{1-\frac u2}}{(s-2)(t-2)}\Big)\Bigg]\,,
}
where $H_n$ is a harmonic number.

 \section{Perturbative string coupling}
\label{pertString}

We will now consider the expansion of the correlator at large $N$ and $\lambda$, which according to the AdS/CFT dictionary \eqref{adscft} corresponds to small string length $\ell_s$ and string coupling $g_s$. This expansion is most easily expressed in terms of the Mellin amplitude $M(s,t) $ in \eqref{mellin}. The advantage of Mellin space is that poles of $M(s, t)$ in $s,t,u$ correspond to single-trace exchange Witten diagrams while polynomials in $s,t,u$ give contact diagrams. At tree level, all corrections to double trace CFT data come the gamma functions in \eqref{mellin} but this is no longer true at loop level. The Mellin amplitude is also constrained by crossing symmetry
\es{crossM}{
M^{ABCD}(s,t)=M^{BACD}(s,u)=M^{CBAD}(u,t)\,,
}
and by the flat space limit \cite{Penedones:2010ue} 
\es{flatGen}{
\mathcal{A}^{ABCD}(s,t)=\lim_{L\to\infty} 8\pi^4 L^4\int\frac{d\beta}{2\pi i}\frac{e^\beta}{\beta^4}\frac{L^4(u+s/w)^2}{16}M^{ABCD}\Big(\frac{L^2}{2\beta}s,\frac{L^2}{2\beta}t\Big)\,,
}
where $\mathcal{A}(s,t)$ is the S-matrix in the 8d flat limit of the AdS$_5\times S^3$ holographic correlator, we included the factor $(u+s/w)^2/4$ that relates the Mellin transform of the reduced correlator $\mathcal{G}$ to that of the full correlator $G$ in \eqref{ward4dN2}, and we suppressed the overall polarization dependence. The overall factor of $L^4$ is due to the factor of $S^3$ in the full AdS$_5\times S^3$ spacetime. The constraints of the pole structure, crossing, and the flat limit fix the Mellin amplitude to take the form \cite{Alday:2021odx,Alday:2021ajh} 
\es{melGen}{
&M = \frac{1}{k}\Big[M_{F^2}+\lambda^{-1}\sum_{i=1}^3 b^i_{F^4} M^i_0+\lambda^{-\frac32}\Big(\sum_{i=1}^3 b^i_{D^2F^4} M^i_0+\sum_{i=1}^2 {\tilde b}^i_{D^2F^4} M^i_1\Big)+O(\lambda^{-2})\Big]\\
&\qquad+ \frac{1}{c}M_{R} +\frac{1}{k^2}\Big[M_{F^2|F^2}+\sum_{i=1}^3 b^i_{0} M^i_0+\log\lambda\sum_{i=1}^3 b^i_\text{log} M^i_0+O(\lambda^{-1})\Big]+O(N^{-3})\,.
}
Here, the tree gluon exchange term takes the form\footnote{\label{csetc}Our structures are related to those of \cite{Alday:2021odx,Alday:2021ajh} as $\mathtt{c}_s=\mathtt{t}_4-\mathtt{t}_5$, $\mathtt{c}_t=\mathtt{t}_5$, $\mathtt{c}_u=-\mathtt{t}_4$, $\mathtt{d}_s=\mathtt{t}_1$, $\mathtt{d}_t=\mathtt{t}_2$, $\mathtt{d}_u=\mathtt{t}_3$, our coefficients are related as $(C_{2,2,2})^2=2/k, (C_{2,2,g})^2=1/(6c)$, and we fixed a typo in the graviton exchange term coefficient.}
\es{treeEx}{
M_{F^2}=\frac{8\mathtt{ t}_5-8\mathtt{ t}_4}{(s-2)(4-s-t)}+\frac{8\mathtt{ t}_5}{(t-2)(4-s-t)}\,,
}
where the overall coefficient is fixed in terms of $k$ by conformal Ward identities \cite{Osborn:1993cr}. The tree level exchange graviton amplitude has not yet been computed, as it requires an infinite sum over graviton KK modes that appear in long multiplets (see \cite{Chester:2023qwo} for discussion in a related case).\footnote{An expression for $M_R$ was given in \cite{Alday:2021ajh}, but this only took into account the leading KK mode, and does not have the expected scaling at large $s,t$. These long multiplet graviton modes can be identified by decomposing $\mathcal{N}=4$ half-BPS operators of dimension $p$ to $\mathcal{N}=2$, where for even $p$ one finds that a long multiplet appears in the decomposition.} The contact terms due to higher derivative corrections $F^4$ and $D^2F^4$ to super-Yang-Mills are down by powers of $\lambda$ at genus-zero, and then have higher genus corrections up by powers of $\lambda/N$ as given by the dictionary \eqref{adscft}. Their Mellin amplitudes are polynomials in $s,t$ that take the form
\es{polyM}{
M^1_0&=\mathtt{t}_1+\mathtt{t}_2+\mathtt{t}_3\,,\quad M^2_0=\mathtt{t}_6+\frac13\mathtt{t}_4-\frac23\mathtt{t}_5\,,\quad M^3_0=\mathtt{t}_7\,,\\ 
M^1_1&=s\mathtt{t}_1+u\mathtt{t}_2+t\mathtt{t}_3\,,\quad M^2_1=(t-2)\mathtt{t}_4+(u-2)\mathtt{t}_5\,,\\ 
}
where the overall genus-zero coefficients $b_{F^4}^i$, $b_{D^2F^4}^i$, and  $\tilde b_{D^2F^4}^i$ in \eqref{melGen} remain unfixed. Note that $F^4$ also has a genus-one correction that is indistinguishable from the 1-loop contact term ambiguity with coefficient $b_{0}^i$. Finally, the 1-loop gluon amplitude was computed in \cite{Alday:2021ajh} in terms of tree level data using the AdS unitarity cut method to get\cite{Aharony:2016dwx} \footnote{\label{dstetc}Our structures are related to those of \cite{Alday:2021ajh} as $\mathtt{d}_{st}=2\,[2\,(\mathtt{t}_1+\mathtt{t}_2+\mathtt{t}_3)-\mathtt{t}_4+2\mathtt{t}_5]$,  $\mathtt{d}_{su}=2\,[2\,(\mathtt{t}_1+\mathtt{t}_2+\mathtt{t}_3)+2\mathtt{t}_4-\mathtt{t}_5]$, $\mathtt{d}_{tu}=2\,[2\,(\mathtt{t}_1+\mathtt{t}_2+\mathtt{t}_3)-\mathtt{t}_4-\mathtt{t}_5]$. We chose the regularization ambiguity $a$ in \cite{Alday:2021ajh} to be $a=-\frac{\gamma}{9}$ for future convenience, which was also found to be natural in position space \cite{Huang:2023oxf}.}
\es{loopM}{
&M_{F^2|F^2}=72(2(\mathtt{t}_1+\mathtt{t}_2+\mathtt{t}_3)-\mathtt{t}_4+2\mathtt{t}_5)\mathcal{B}({s,t})\\
&\quad+72(2(\mathtt{t}_1+\mathtt{t}_2+\mathtt{t}_3)+2\mathtt{t}_4-\mathtt{t}_5)\mathcal{B}({s,u})+72(2(\mathtt{t}_1+\mathtt{t}_2+\mathtt{t}_3)-\mathtt{t}_4-\mathtt{t}_5)\mathcal{B}({t,u})\,,
}
where the $\mathcal{B}(s, t)$ have poles at the same locations as the gamma functions and admit the resummed expressions 
\es{resummed}{
\mathcal{B}({s,t}) &= R_0(s, t) \left [ \psi^{(1)}(2 - \tfrac{s}{2}) + \psi^{(1)}(2 - \tfrac{t}{2}) - \left ( \psi(2 - \tfrac{s}{2}) - \psi(2 - \tfrac{t}{2}) \right )^2  \right ] \\
&+ R_1(s, t)\psi^{(0)}(2 - \tfrac{s}{2}) + R_1(t, s)\psi^{(0)}(2 - \tfrac{t}{2}) + \pi^2 R_2(s, t) + \frac{4}{27(s + t - 8)} -\frac{\gamma}{9}\,.
}
Here $\gamma$ is the Euler constant and we have the coefficient functions 
\es{Rs}{
R_0(s, t) &= \frac{3s^2t - 8s^2 + 3st^2 - 32st + 60s - 8t^2 + 60t - 96}{54(s + t - 8)(s + t - 6)(s + t - 4)}  \\
R_1(s, t) &=- \frac{3s^2 + 3st - 26s - 10t + 48}{27(s + t - 8)(s + t - 4)}, \quad R_2(s, t) = -R_0(s, t)\,.
}
 At large $s,t$ this gives \cite{Alday:2021ajh}
\es{boxFin2}{
\mathcal{B}({s,t})&\sim -\frac{s t \log ^2\left(\frac{-s}{-t}\right)}{18(s+t)^2}-\frac{s \log (-s)+t \log (-t)}{9(s+t)}-\frac{\pi^2 s t}{18(s+t)^2}-\frac{1}{9}(\gamma-\log 2)\,.
}
The AdS unitarity method fixed this 1-loop amplitude up to a contact term ambiguity with coefficients $b_{0}^i$, and we also expect a $\log\lambda$ term with coefficients $b^i_\text{log}$ that regularizes the logarithmic divergence of the 1-loop term. We will now fix the various $b$ coefficients using localization and then the known flat space limit.

\subsection{Constraints from supersymmetric localization}
\label{perLoc}

We will now constrain the holographic correlator using the mass deformed free energy $F(\mu_i)$, whose derivatives are related to the integrated correlator as in Section \ref{intCon}. The partition function $Z(\mu_i)\equiv e^{-F(\mu_i)}$ is computed using localization in terms of a matrix model integral \cite{Pestun:2016zxk} as\footnote{When we discuss non-perturbative string coupling later (and everywhere in Appendix \ref{tseytlin}), we will switch to a different notation in which $Z_\text{inst}(X, \mu_i, \tau_\text{UV})$ refers to only part of the instanton partition function, and there is an extra instanton factor called $Z_\text{extra}$.}
\es{Z}{
Z(\mu_i)=\int [dX]\,e^{{-\frac{8\pi^2}{g_\text{YM}^2}}\,\text{tr}X^2}\,|Z_{\text{1-loop}}(X,\mu_i)|^2\,|Z_{\text{inst}}(X,\mu_i,\tau_\text{UV})|^2\,,
}
where $X$ are $2N\times 2N$ matrices in the Lie algebra $\mathfrak{sp}(2N)$ with eigenvalues $\{\pm x_1,\ldots,\pm x_N\}$, and the Vandermonde measure is
\es{Van}{
[dX]=\frac{1}{N!}\prod_{i=1}^N dx_n\,x^2_n\prod_{1\leq n<m\leq N}(x^2_n-x^2_m)^2\,.
}
We are interested in the expansion for large $N$ and large $\lambda$, so we can neglect the instanton contribution, while the 1-loop term is
\es{Zloop}{
|Z_{\text{1-loop}}(X,\mu_i)|^2\equiv e^{-S_{\mathrm{int}}(X,\mu_i)}=\prod_{n=1}^N\frac{H(2x_n)^2}{\prod_{i=1}^4 H(x_n+\mu_i)\,H(x_n-\mu_i)}\,,
}
and $H(x)=e^{-(1+\gamma)x^2}G(1+i x)G(1-ix)$ is written in terms of the Barnes G function. The 1-loop term can  can be written as an interaction term $S_{\mathrm{int}}(X,\mu_i)$ in the action, which makes the matrix model interacting even for $\mu_i=0$. As shown in \cite{Beccaria:2021ism,Beccaria:2022kxy}, since the interaction terms are all single-trace deformations of the free gaussian matrix model, this allows $F(\mu_i)$ to be computed to all orders in $1/N$ at fixed $\lambda_\text{UV}=g_\text{YM}^2 N$ using so-called Toda equations. In fact, \cite{Beccaria:2022kxy} only considered the case of a single mass, but in Appendix \ref{tseytlin} we show that the general mass case is a straightforward generalization that gives the all orders result for the mass derivatives
\es{largeLamF}{
-\partial^4_{\mu_1}F|_{\mu=0}&=\frac{32\pi^2}{\lambda}N+6\log\lambda - 16 \log 2 + 3f(N)+O(e^{-N},e^{-\lambda})\,,\\
-\partial^2_{\mu_1}\partial^2_{\mu_2}F|_{\mu=0}&=2\log\lambda+f(N)+O(e^{-N},e^{-\lambda})\,,\\
-\partial_{\mu_1}\partial_{\mu_2}\partial_{\mu_3}\partial_{\mu_4}F|_{\mu=0}&=O(e^{-N},e^{-\lambda})\,,
}
where here we used the IR 't Hooft coupling $\lambda$ \eqref{AdSV} and we define
\es{tF}{
f(N)&=\frac{16}{3}-4 \zeta (3)+4 \gamma -4 \log (4 \pi )-2\log(2N)+2\psi(2 N+5/2)\\
&+(N+1/4)\psi^{(1)}(N+{5}/{4})+(N+3/4)\psi^{(1)}(N+{7}/{4})\\
&= \frac{22}{3}  -4 \zeta (3)+4 \gamma -4 \log (4 \pi )+ \frac{1}{N} - \frac{7}{48N^2} - \frac{1}{48N^3} + \frac{67}{2560N^4} + O(N^{-5})\,.
}
We can then impose these values on the holographic correlator \eqref{melGen} using the three integrated constraints \eqref{intNew}. We make use of the Mellin space integrals \eqref{Imel} of the functions\footnote{The $SO(8)$ structures of the gluon exchange term vanish in the integrated constraint, so we need not consider its integral.}
\es{Is}{
I[1]=\frac{1}{24}\,,\qquad \qquad I[s]=I[t]=I[u]=\frac{1}{12}\,,
}
which we computed to high precision and matched to these analytic values.\footnote{In Appendix \ref{integration} we also show how to derive some of these integrals analytically.} We then apply \eqref{intNew} to the Mellin amplitudes in \eqref{melGen} to constrain three coefficients at each order in $1/N$ and $1/\lambda$, which fixes the $b$ to 
\es{bStrong}{
b^1_{\log}&=48\,,\qquad b^2_{F^4}=96{\pi^2}\,, \qquad b^2_{0}=-192\log2\,,\qquad b^1_{D^2F^4}=-2\tilde b^1_{D^2F^4}\,,\\
 b^2_{\log}&=b^3_{\log}=b^1_{F^4}=b^3_{F^4}=b^2_{D^2F^4}=b^3_{D^2F^4}=b_0^3=0\,,
}
as well as $\tilde b^2_{D^2F^4}$ which remains unfixed. The localization constraints also imply that all higher genus corrections to $F^4$ vanish as expected. Note that in principle we could also fix $b_0^1$ using the localization constraint, which is affected by both the graviton exchange $M_R$ and 1-loop term $M_{F^2|F^2}$, but the explicit expression for $M_R$ is not yet available.

\subsection{Constraints from flat space limit}
\label{flat}

We will now compare the flat space limit \eqref{flatGen} of the AdS holographic correlator to the flat space Type IIB amplitude $\mathcal{A}(s,t)$ with D7 branes in the small $g_s$ expansion:
\es{Agen}{
\mathcal{A}(s,t)=(u+s/w)^2\Big[g_s \mathcal{A}_V(\ell_s)+g_s^2\big(\ell_s^8[\cA_{F^2|F^2}+\cA_{R}+\cA_0]+\ell_s^8\log[\ell_s^2]\cA_\text{log}+O(\ell_s^{10})\big)+O(g_s^3)\Big]\,,
}
 where $s,t,u=-s-t$ are Mandelstam variables\footnote{Our conventions are $t\leftrightarrow u$ compared to the standard conventions in \cite{Polchinski:1998rr} so as to match our Mellin space conventions.} and we suppress the overall polarization dependence as in \eqref{flatGen}. The genus-zero term $\mathcal{A}_V$ can be written at finite $\ell_s$ in terms of the Veneziano amplitude \eqref{V}, which we can expand in string length to get SYM plus higher derivative corrections:
\es{V2}{
\mathcal{A}_V=&-32\pi^5{\ell_s^4}\left[\frac{t(\mathtt{t}_4-\mathtt{t}_5)-s\mathtt{t}_5}{stu}+\ell_s^4\zeta(2)(2\mathtt{t}_5-\mathtt{t}_4-3\mathtt{t}_6)\right.\\
&\left.\qquad\qquad\qquad\qquad\quad+\ell_s^6\zeta(3)(t(\mathtt{t}_4-\mathtt{t}_5)-s\mathtt{t}_5)+O(\ell_s^8)\right]\,,
}
where we show the $F^4$ and $D^2F^4$ corrections. Note that at leading and $D^2F^4$ orders the $\mathtt{t}_6$ structure vanishes because $s+t+u=0$, but this structure appears at all other orders including $F^4$ shown here. We normalized this amplitude so that its leading term arises from 8d SYM with the standard action
\es{8dS}{
S=\frac{1}{4\mathtt{g}_{8d}^2}\int d^8x F^A_{\mu\nu}F^{A,\mu\nu}\,,\qquad \mathtt{g}_{8d}^2={64\pi^5 g_s \ell_s^4}\,,
}
where $\mu,\nu=1,\dots8$ are 8d spacetime indices, $A$ is an adjoint index of the $SO(8)$ gauge group, and as shown in Appendix \ref{8damp} our brane construction fixes the value of $\mathtt{g}_{8d}^2$. This action can be used to compute the non-analytic in $s,t$ part of the 1-loop correction, which as shown in Appendix \ref{8damp} takes the form 
\es{Aloop}{
&\mathcal{A}_{F^2|F^2} ={2^{12}}\pi^{10}\big[(2(\mathtt{t}_1+\mathtt{t}_2+\mathtt{t}_3)-\mathtt{t}_4+2\mathtt{t}_5)\mathcal{A}_{st}\\
&+(2(\mathtt{t}_1+\mathtt{t}_2+\mathtt{t}_3)+2\mathtt{t}_4-\mathtt{t}_5)\mathcal{A}_{su}+ (2(\mathtt{t}_1+\mathtt{t}_2+\mathtt{t}_3)-\mathtt{t}_4-\mathtt{t}_5)\mathcal{A}_{tu}\big]\,,\\
}
where $\mathcal{A}_{st}$ is proportional to a regularized 8d box diagram
\es{boxFin}{
\mathcal{A}_{st}&=-\frac{1}{3072\pi^4}\left(f_{\text{box}}^{\text{8d}}(s,t)-2\gamma\right)\,,
}
where
\es{fbox8d}{
f_{\text{box}}^{\text{8d}}(s,t)&=\frac{s t \log ^2\left(\frac{-s}{-t}\right)}{(s+t)^2}+\frac{2(s \log (-s)+t \log (-t))}{s+t}+\frac{\pi^2 s t}{(s+t)^2}\,,
}
with $s,t<0$ and the regularization generates analytic terms that we chose to be proportional to the large $s,t$ limit of \eqref{loopM}. The ambiguity in these analytic terms can be pushed into $\cA_0(s,t)$, which also includes the genus-one correction to $F^4$. 

The term $\mathcal{A}_R$ in \eqref{Agen} refers to the exchange of a 10d graviton between 8d gluons, where note that momentum is only conserved along the brane so although it is a tree amplitude there is still a loop and from \eqref{gravitonexch} we find
\es{AR}{
\cA_R=-16\pi^6\,\left[\mathtt{t}_1\,\log (-s)+\mathtt{t}_2\,\log (-t)+\mathtt{t}_3\,\log (-u)\right]\,,
}
where once again we have pushed the ambiguous analytic terms into $\cA_0(s,t)$. 

Lastly, since $s,t$ are dimensionful the $\log(-s)$ and $\log(-t)$ terms arising from the one-loop amplitude \eqref{Aloop} and from the graviton exchange amplitude \eqref{AR} must come with $\log(\ell_s^2)$ terms 
\es{logFlat}{
\cA_\text{log}=-32{\pi^6}(\mathtt{t}_1+\mathtt{t}_2+\mathtt{t}_3)\,,
}
which regularizes the logarithmic divergences that occur in the tree graviton and 1-loop gluon amplitudes.

We can now use the flat space limit formula \eqref{flatGen} to compare the AdS correlator $M(s,t)$ to the flat amplitude \eqref{Agen} after converting $N,\lambda$ to $\ell_s,g_s$ using \eqref{adscft}, where the explicit relation between $g_s$ and $\lambda$ is $\lambda=8\pi N g_s$. At genus-zero, we find a precise match between the $F^4$ coefficients computed from localization in \eqref{bStrong} and the leading correction to the Veneziano amplitude \eqref{V}. For $D^2F^4$ we can combine the known flat space limit with the localization constraints to completely fix the AdS coefficients to 
\es{D2F4pert}{
\tilde{b}^2_{D^2F^4}=-1536\zeta(3)\,,\qquad b^1_{D^2F^4}=\tilde b^1_{D^2F^4}=0\,.
}
The genus-zero AdS amplitude can then be written analogously to the flat amplitude \eqref{V} as \eqref{AdSV}, as discussed in the introduction. 

At genus-one, we find a precise match between the  $\log\lambda$ computed from localization in \eqref{bStrong} and the $\log\ell_s^2$ that regularizes the 1-loop flat space amplitude, as well as between the non-analytic 1-loop Mellin amplitude in \eqref{loopM} and the flat space 1-loop amplitude in \eqref{boxFin}. 

Finally, while we know the flat limit of $M_R$, the mellin amplitude itself has not yet been computed so we cannot yet compare it.

 \section{Non-perturbative string coupling}
\label{NPString}

We will now consider the expansion of the correlator at large $N$ and finite $\tau$, which according to the AdS/CFT dictionary \eqref{adscft} corresponds to small Planck length $\ell_P\equiv g_s^{1/4}\ell_s$ and finite string coupling $\tau_s$. 
The large $N$ holographic correlator as constrained by the analytic bootstrap can be obtained from the large $N$, large $\lambda$ expansion \eqref{melGen} by simply writing $\lambda=8\pi N{g_s}$ and re-expanding in $N$. We get
\es{melGen2}{
M &= \frac{M_{F^2}}{4N}+\frac{1}{N^2}\Big[2M_R+\frac{M_{F^2|F^2}}{16}+\sum_{i=1}^3 \frac{a^i_{0} M^i_0}{16}+\sum_{i=1}^3 a^i_{F^4}(\tau) M^i_0+\log N\sum_{i=1}^3 a^i_{\log }(\tau) M^i_0\Big]\\
&\qquad\qquad\qquad\qquad+\frac{\sum_{i=1}^3 a^i_{D^2F^4}(\tau) M^i_0+\sum_{i=1}^2 {\tilde a}^i_{D^2F^4}(\tau) M^i_1}{N^{\frac52}}+O(N^{-3})\,,
}
where the Mellin amplitudes were defined, and the 1-loop term is now naturally regularized by $\log N\sim\log\ell_P$. We divide the other $O(N^{-2})$ terms into the $\tau$ dependent contributions to $F^4$, denoted as $a_{F^4}^i(\tau)$, and the $\tau$-independent terms $a_0^i$ that are the same as $b_0^i$ in the weak coupling limit. The coefficients $a(\tau)$ (except $a_0^i$) are now functions of finite $\tau$, which we will fix in the next couple sections using localization and then the known flat space limit. 

\subsection{Constraints from supersymmetric localization}
\label{perLoc2}

At large $N$ and finite $\tau$ we can divide $F(\mu_i)$ into perturbative in $\tau_2$ terms for $\tau\equiv\tau_1+i\tau_2$, and non-perturbative in $\tau$ terms that will come from instantons. We can obtain the perturbative terms from the large $N$ large $\lambda$ results \eqref{largeLamF} by simply setting $\lambda=\frac{ 8\pi N}{\tau_2}$ to get
\es{largeNpert}{
\mathcal{F}_v &= -8 \log \tau_2 + 8\log [{8\pi N}]+4f(N)+O(e^{-N},e^{i\tau}) \\
\mathcal{F}_c = \mathcal{F}_s &= 4\pi \tau_2-16\log2  -8 \log \tau_2 + 8\log [{8\pi N}]+4f(N)+O(e^{-N},e^{i\tau})\,, \\
}
where we reorganized the mass derivatives as \eqref{Fcsv} to better see the relation between $SL(2,\mathbb{Z})$ duality and $SO(8)$ triality. In paricular, at finite $\tau$ duality acts on the $SO(8)$ triality frame as \cite{Seiberg:1994aj,Sen:1996vd}
\es{dualTrial}{
&S:\qquad\tau\to-1/\tau\qquad \Leftrightarrow \qquad{\bf 35_v} \leftrightarrow {\bf 35_c} \qquad\Leftrightarrow   \qquad{\mathcal{F}_{\bf v}} \leftrightarrow {\mathcal{F}_{\bf c}}\,,\\
&T:\qquad\tau\to\tau+1\qquad \Leftrightarrow \qquad{\bf 35_c} \leftrightarrow {\bf 35_s}\qquad\Leftrightarrow   \qquad{\mathcal{F}_{\bf c}} \leftrightarrow {\mathcal{F}_{\bf s}}\,. \\
}
To compute the $O(e^{i\tau})$ terms we must consider the contribution to the matrix model \eqref{Z} from the instanton term $Z_{\text{inst}}(X,\mu_i,\tau_\text{UV})$, which can be expanded as
\es{Zinst}{
Z_{\text{inst}}(X,\mu_i,\tau_\text{UV})=\sum_{k=0}^\infty e^{2\pi i \tau_\text{UV}} Z_k(X,\mu_i)=\sum_{k=0}^\infty e^{\pi i \tau} Z'_k(X,\mu_i)\,,
}
where in the second equality we used the UV-IR relation \eqref{UVtoIR} to write the instanton expansion in terms of coefficients $ Z'_k(X,\mu_i)$ that are linear combinations of the $Z_k(X,\mu_i)$. While these coefficients can be directly computed in 4d using the ADHM construction for $U(N)$ gauge group \cite{Nekrasov:2002qd,Nekrasov:2003rj}, for other gauge groups such as $USp(2N)$ the naive 4d formulation is singular so instead one must perform the computation in 5d and then dimensionally reduce to 4d. This was first done in \cite{Nekrasov:2002qd,Nekrasov:2003rj,Shadchin:2005mx} to get the coefficients
\es{Zdetail}{
Z_k(X,\mu_i)=\int \prod_{I=1}^n\frac{d\phi_I}{2\pi i}z_k^{\text{anti}}(X,\phi)z_k^{\text{vec}}(X,\phi)\prod_{i=1}^4z_k^{\text{hyper}(i)}(X,\phi,\mu_i)\,,
}
where in Appendix \ref{tseytlin} we give the explicit formulae as well as the contour prescription. 

However, this prescription for the instanton contributions cannot be complete for three reasons. Firstly, for the $USp(2)$ theory the antisymmetric multiplet becomes a singlet, so naively we would expect that it does not contribute, making the theory equivalent to $SU(2)$ SQCD. As shown in \cite{Hollands:2010xa}, the localization formula for $USp(2)$ and $SU(2)$ can indeed be made equivalent after identifying the IR $\tau$ in each theory, but only if the contribution to \eqref{Zdetail} from the antisymmetric multiplet is removed by hand. The second problem with \eqref{Zdetail} is that if we consider this expression for general equivariant parameters $\epsilon_1,\epsilon_2$, then the UV to IR relation in \eqref{UVtoIR} (\textit{i.e.} the Seiberg-Witten curve) should be recovered from\footnote{It is common to see \eqref{prepot} written with $2\pi i \tau$ on the left hand side. We instead define $\tau$ to be twice as large so that it has the standard transformation properties under $SL(2, \mathbb{Z})$. This rescaling arises from the difference between $SU(N)$ and $USp(2N)$ in the Killing form that appears in the classical contribution to $Z$ \cite{Alday:2021vfb}.}
\es{prepot}{
\pi i \tau =2\pi i \tau_\text{UV} -\frac{1}{2} \frac{\partial^2}{\partial x^2} \lim_{\epsilon_{1,2} \to 0} \epsilon_1 \epsilon_2 \log Z_\text{inst}(X, 0, \tau_\text{UV})
}
as explained {\it e.g.} in \cite{Hollands:2010xa}, but this is not the case with \eqref{Zdetail}. The third problem is that the mass derivatives in \eqref{Fcsv} must be permuted by triality as in \eqref{dualTrial}. While this property is only precise after including all of the infinitely many instantons, it should be visible approximately even with just a finite number of instantons, as was checked numerically for $SU(2)\cong USp(2)$ in \cite{Chester:2022sqb}. All these issues show there must be an extra contribution to the instanton terms that is not captured by \eqref{Zdetail}.

This subtlety was first noticed in a related 5d context in \cite{Kim:2012gu,Hwang:2014uwa}. In 5d one can consider a family of $\mathcal{N}=1$ CFTs with $USp(2N)$ gauge group, 1 antisymmetric hypermultiplet, and $0\leq N_f\leq7$ fundamental hypermultiplets \cite{Seiberg:1996bd}. These theories are expected to have enhanced $E_{N_f+1}$ global symmetry. The $S^4\times S^1$ superconformal index was computed using localization from a certain ADHM gauged quantum mechanics. However, this quantum mechanics contains extra degrees of freedom $Z^{\text{5d}}_\text{extra}$, which are $N$-independent, and must be explicitly divided out to get the correct index with the expected enhanced symmetry. For $N_f=4$, the calculation of the $S^4\times S^1$ index in this 5d CFT reduces to the $S^4$ free energy in our 4d CFT upon shrinking the $S^1$. The contribution of the quantum mechanics gives the $Z_{\text{inst}}(X,\mu_i,\tau)$ terms in \eqref{Zdetail} as earlier shown by \cite{Nekrasov:2002qd,Nekrasov:2003rj,Shadchin:2005mx}, and we expect a new contribution $Z_\text{extra}$ from the 4d version of $Z^{\text{5d}}_\text{extra}$, such that we must redefine the full partition function \eqref{Z} as
\es{Znew}{
Z(\mu_i)&=\int [dX]\,e^{{-2\pi \Im(\tau_\text{UV})}\,\text{tr}X^2}\,|Z_{\text{1-loop}}(X,\mu_i)|^2\,|Z_{\text{inst}}(X,\mu_i,\tau_\text{UV})|^2\\
&\to \int [dX]\,\left(\frac{e^{{-\pi \Im(\tau)}}}{16}\right)^{\text{tr}X^2}\,|Z_{\text{1-loop}}(X,\mu_i)|^2   \frac{ |Z_{\text{inst}}(X,\mu_i,\tau)|^2}{|Z_\text{extra}(X,\mu_i,\tau)|^2} \,.
}
Note that in the second line we defined $Z_\text{extra}$ such that the classical term is written terms of the IR quantity $\tau$ according to \eqref{UVtoIR}, which changes the dependence of $Z_\text{extra}$ on the eigenvalues $X$.

\begin{figure}[h]
\centering
\includegraphics[scale=0.5]{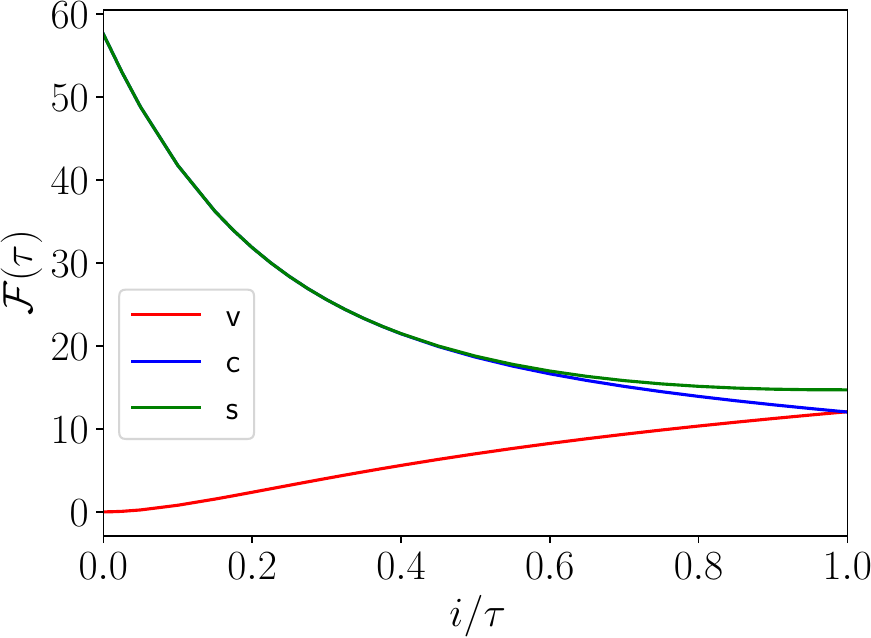} \quad \includegraphics[scale=0.5]{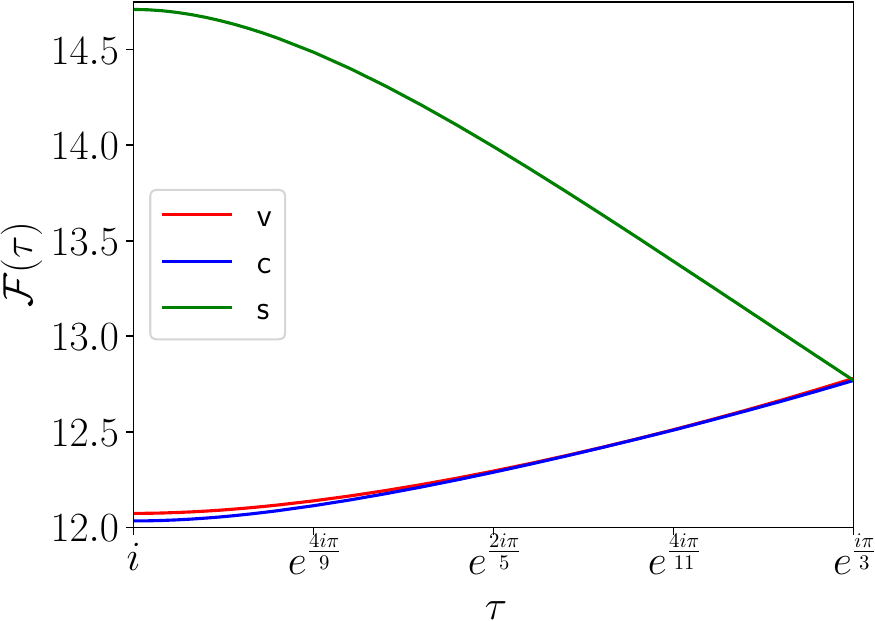}
\caption{The localization inputs for $\mathcal{F}_{\bf v}$, $\mathcal{F}_{\bf c}$, and $\mathcal{F}_{\bf s}$ in the $USp(4)$ theory as a function of $i/\tau $ from the free point $\tau=i\infty$ until the $S$-duality invariant point $\tau=i$ with $\Re(\tau)=0$ (\textbf{left}), and as a function of $\tau$ along the $S$-duality invariant arc with $|\tau|=1$ from $\tau=i$ to the $T$-duality invariant point $\tau=e^{i \pi /3}$ (\textbf{right}). These functions were computed up to 2-instanton order using \eqref{ZextraFinal}, and show some small errors (e.g. $\mathcal{F}_{\bf v}\neq \mathcal{F}_{\bf c}$ at $\tau=i$ in the right hand plot). The analogous plots for $USp(2)$ are shown in \cite{Chester:2022sqb}, where we could go to much higher instanton order due to the equivalence to $SU(2)$ SQCD. }
\label{usp4check}
\end{figure}

We then compute $Z_\text{extra}$ directly in 4d from two assumptions.\footnote{While it would be ideal to simply derive $Z_\text{extra}$ from a dimensional reduction of $Z^{\text{5d}}_\text{extra}$, we do not yet know how to do this. See Appendix \ref{tseytlin} for further discussion.} Firstly, we consider the $USp(2)$ theory with $Z(\mu_i)$ defined as in \eqref{Znew}, and demand that it be equivalent to $SU(2)$ SQCD as expected, after identifying the IR $\tau$ in each case. As discussed above, this gives a nontrivial value of $Z_\text{extra}$, because $Z_{\text{inst}}$ has unexpected contributions from the antisymmetric hypermultiplet. We perform this comparison to 2-instanton order in Appendix \ref{tseytlin} to get\footnote{For generality, we consider also the $N_f<4$ asymptotically free gauge theories.} 
\es{ZextraFinal}{
Z_\text{extra}(X,\mu_i,\tau)&=1-8e^{\pi i \tau}\prod_{i=1}^4\mu_i+e^{2\pi i \tau}\,\left[\frac{109}{8}-3\sum_{i=1}^4\mu_i^2-2\sum_{i<j\leq 4}\mu_i^2\mu_j^2+32\prod_{i=1}^4\mu_i^2\right]+O(\mu^6,e^{3\pi i \tau})\,.
}
Crucially, to this order we find that $Z_\text{extra}(X,\mu_i,\tau)$ is independent of the eigenvalues $X$ (so that we will henceforth call it $Z_\text{extra}(\mu_i,\tau)$), where at 2-instanton order this is affected by the definition of the classical term in \eqref{Znew}. Arranging for this to happen at 3-instanton order appears to be more difficult but substantial evidence for our results will be available without this. Our second assumption is that we expect $Z_\text{extra}$ to hold for all $USp(2N)$, as was the case for $Z^{\text{5d}}_\text{extra}$. As a consistency check, we then numerically compute mass derivatives of $Z(\mu_i)$ with $Z_\text{extra}$ given in \eqref{ZextraFinal} for both $USp(2)$ and $USp(4)$. For $USp(2)$, which allows the antisymmetric hypermultiplet to be removed trivially, we have been able to do this to 4-instanton order. This leads to numerical $\mathcal{F}_{\bf v}$, $\mathcal{F}_{\bf c}$ and $\mathcal{F}_{\bf s}$ functions which are essentially identical to the 8-instanton ones plotted in \cite{Chester:2022sqb} so we do not show them here. For $USp(4)$, our 2-instanton calculation is enough to show that the expected $SL(2,\mathbb{Z})$ triality relation in \eqref{dualTrial} significantly improves due to the inclusion of $Z_\text{extra}$.\footnote{In detail, at the fixed point $\tau=e^{i\pi/3}$, $\mathcal{F}_{\textbf{c}, \textbf{s}, \textbf{v}}$ start off within $1.8 \%$ of each other at $k = 0$ instantons. This error shrinks to $0.1 \%$ at $k = 2$. Without $Z_\text{extra}$, the error would instead grow.} 
The agreement is visible in Figure \ref{usp4check} especially at the $\tau = i$ and $\tau = e^{i \pi / 3}$ fixed points of the duality transformation.

Now that we have a complete expression for $Z(\mu_i)$ for any $N$, we can finally consider the large $N$ finite $\tau$ limit of the instanton contributions. As shown in Appendix \ref{tseytlin}, the contributions from the usual instanton terms $Z_{\text{inst}}(X,\mu_i,\tau)$ are exponentially small at large $N$ and finite $\tau$ expansion, so that instanton corrections come only from $Z_\text{extra}(\mu_i,\tau)$ and only appear at order $O(N^0)$. After taking mass derivatives and combining with the perturbative corrections in \eqref{largeNpert} we find results consistent to 2-instanton order with
\es{Ffinal}{
\mathcal{F}_{\bf v}&=-8\log[\tau_2 |\theta_3(\tau)\theta_4(\tau)|^2]+8\log[8\pi N]+4f(N)+O(e^{-N})\,,\\
\mathcal{F}_{\bf c}&=-8\log[\tau_2 |\theta_2(\tau)\theta_3(\tau)|^2]+8\log[8\pi N]+4f(N)+O(e^{-N})\,,\\
\mathcal{F}_{\bf s}&=-8\log[\tau_2 |\theta_2(\tau)\theta_4(\tau)|^2]+8\log[8\pi N]+4f(N)+O(e^{-N})\,,\\
}
where these Jacobi theta functions are defined as
\es{jacobi}{
\theta_2(\tau) &= 2e^{\frac{\pi i\tau}{4}} \prod_{n = 1}^\infty (1 - e^{2n\pi i\tau})(1 + e^{2n\pi i\tau})^2  \\
\theta_3(\tau) &= \prod_{n = 1}^\infty (1 - e^{2n\pi i\tau}) \left ( 1 + e^{(2n-1)\pi i\tau} \right )^2  \\
\theta_4(\tau) &= \prod_{n = 1}^\infty (1 - e^{2n\pi i\tau}) \left ( 1 - e^{(2n-1)\pi i\tau} \right )^2\,.
}
These functions transform under $SL(2,\mathbb{Z})$ as
\es{Jtrans}{
\theta_2 \left ( -{1}/{\tau} \right ) &= \sqrt{-i \tau} \theta_4(\tau) \,,\qquad \theta_2(\tau + 1) = e^{\frac{\pi i}{4}} \theta_2(\tau)  \\
\theta_3 \left ( -{1}/{\tau} \right ) &= \sqrt{-i \tau} \theta_3(\tau) \,,\qquad \theta_3(\tau + 1) = \theta_4(\tau)  \\
\theta_4 \left ( -{1}/{\tau} \right ) &= \sqrt{-i \tau} \theta_2(\tau)  \,,\qquad \theta_4(\tau + 1) = \theta_3(\tau) \,,
}
so that the mass derivatives in \eqref{Ffinal} are permuted by duality as expected.

We can then impose these values on the holographic correlator \eqref{melGen2} using the three integrated constraints \eqref{intOld} to fix three coefficients at each order in $1/N$ to get 
\es{bvStrong}{
a^1_{F^4}&=-3\log\Big[\tau_2 |\theta_3(\tau)\theta_4(\tau)|^2\Big]\,,\\
 a^2_{F^4}&=-3\log\Big[\tau_2 |\theta_2(\tau)\theta_3(\tau)|^2\Big]-3\log\Big[\tau_2|\theta_2(\tau)\theta_4(\tau)|^2\Big]+6\log\Big[\tau_2 |\theta_3(\tau)\theta_4(\tau)|^2\Big]\,,\\
  a^3_{F^4}&=18\log\Big[\tau_2 |\theta_2(\tau)\theta_3(\tau)|^2\Big]-18\log\Big[\tau_2 |\theta_2(\tau)\theta_4(\tau)|^2\Big]\,,\\
a^1_{\log}&=3\,,\qquad a^1_{D^2F^4}=-2\tilde a^1_{D^2F^4}\,,\qquad a^2_{\log}=a^3_{\log}=a^2_{D^2F^4}=a^3_{D^2F^4}=0\,,
}
where $\tilde a^2_{D^2F^4}$ remains unfixed. Note that by definition we only write the $\tau$-dependent terms in $a_{F^4}^i(\tau)$, while all $\tau$-independent terms are included in the $a_0^i$ that we do not fix due to their dependence on the incomplete 1-loop term. The $F^4$ term in the irrep basis is then\footnote{In writing this result using $\eta(\tau)$, we neglect some $\tau$-independent $\log2$ terms.}
\es{F4finT}{
&\sum_{i=1}^3 a^i_{F^4}(\tau) M^i_0=  -180\log[\sqrt{\tau_2}|\eta(\tau)|^2]P_{\bf 1}
- 12\log[\sqrt{\tau_2}|\eta(\tau)|^2]
P_{\bf 300}\\
&+3\Big[  8\log[\sqrt{\tau_2}|\eta(\tau)|^2]-12\log[\sqrt{\tau_2}|\theta_2(\tau)|^2]
\Big]P_{\bf 35_v}\\
&+3\Big[  8\log[\sqrt{\tau_2}|\eta(\tau)|^2]-12\log[\sqrt{\tau_2}|\theta_4(\tau)|^2]
\Big]P_{\bf 35_c}\\
&+3\Big[  8\log[\sqrt{\tau_2}|\eta(\tau)|^2]-12\log[\sqrt{\tau_2}|\theta_3(\tau)|^2]
\Big]P_{\bf 35_s}\,,
}
where $\eta(\tau)$ is the Dedekind eta function 
\es{eta}{
\eta(\tau) = e^{\frac{\pi i \tau}{12}} \prod_{n = 1}^\infty (1 - e^{2n\pi i \tau })\,,\qquad \eta(\tau)^3=\frac12\theta_2(\tau)\theta_3(\tau)\theta_4(\tau)\,,
}
which transforms under $SL(2,\mathbb{Z})$ as
\es{Etrans}{
\eta \left ( -{1}/{\tau} \right ) &= \sqrt{-i \tau} \eta(\tau) \,,\qquad \eta(\tau + 1) = e^{\frac{\pi i}{12}} \eta(\tau) \,.
}
From these $SL(2,\mathbb{Z})$ transformations we see that ${\bf35_{c,s,v} }$ are permuted as expected, while the other irreps are invariant.

\subsection{Constraints from flat space limit}
\label{flat2}

We will now compare the flat space limit \eqref{flatGen} of the AdS holographic correlator to the flat space Type IIB amplitude $\mathcal{A}(s,t)$ at finite $\tau_s$ and large $\ell_P$:
\es{Agen2}{
\mathcal{A}(s,t)&= (u+s/w)^2\Big[\ell_P^4\mathcal{A}_{F^2}+\ell_P^8\big[\cA_{F^4}(\tau_s)+\cA_{F^2|F^2}+\cA_0+\cA_R+\cA_\text{log}\log\ell_P^2\big]\\
&+\ell_P^{10}\cA_{D^2F^4}(\tau_s)+O(\ell_P^{12})\Big]\,.
}
The $\tau_s$-independent terms are by definition the same as in the perturbative string amplitude \eqref{Agen}, except setting $g_s\ell_s^4=\ell_P^4$. For instance, the leading 8d SYM term in the irrep basis is 
\es{tauInd}{
\cA_{F^2}=2^{6}\pi^5\frac{6sP_{\bf1}+3(t-u)P_{\bf28}+2s(P_{\bf35_c}+P_{\bf35_s}+P_{\bf35_v})-sP_{\bf300}}{stu}\,.
}
The $\tau$-dependence of the protected $F^4$ term was computed using duality to heterotic string theory in \cite{Bachas:1997mc,Gutperle:1999xu,Bachas:1997xn,Foerger:1998kw,Bianchi:1998vq,Gava:1999ky,Kiritsis:2000zi}, as well as by computing D(-1) instanton amplitudes in Type IIB string theory in \cite{Gava:1999ky,Billo:2009gc,Billo:2010mg}, and precisely matches the flat limit of the Mellin space $F^4$ \eqref{F4finT} as 
\es{F4flat}{
\cA_{F^4}(\tau_s)=\frac{16\pi^6}{3}\sum_{i=1}^3 a^i_{F^4}(\tau_s) M^i_0\,,
}
where the prefactor comes from the flat limit formula \eqref{flatGen} and the AdS/CFT dictionary \eqref{adscft}. This is a check of AdS/CFT in this context at finite string coupling. Note that the $\ell_s^8\log[\ell_s^2]$ threshold term, as fixed from the 1-loop term in Section \ref{flat}, precisely combines with the $\ell_s^8$ term from the Veneziano amplitude to give the perturbative contributions to $F^4$ at finite $\tau_s$, which is a check on the relative normalization of these terms.

The $D^2F^4$ term is also protected, and so we expect to also be able to compute its coefficient at finite $\tau_s$. As discussed above, we can see from the Veneziano amplitude \eqref{V2} that $\cA_{D^2F^4}(\tau_s)$ has the same triality invariant flavor structure as $\cA_{F^2}$ in \eqref{tauInd}, unlike all other higher derivative corrections including $\cA_{F^4}(\tau_s)$. This implies that $\cA_{D^2F^4}(\tau_s)$ can be written as 
\es{D2F4}{
\cA_{D^2F^4}(\tau_s)=g(\tau_s,\bar\tau_s)(6sP_{\bf1}+3(t-u)P_{\bf28}+2s(P_{\bf35_c}+P_{\bf35_s}+P_{\bf35_v})-sP_{\bf300})\,,
}
where $g(\tau_s,\bar\tau_s)$ is an $SL(2,\mathbb{Z})$ invariant function of complex $\tau_s$. We can then compute $g(\tau_s,\bar\tau_s)$ following the very similar cases of \cite{Wang:2015jna,Lin:2015ixa}, which we briefly sketch here (for more details see the original papers). Since the gravity multiplet contains fluctuations of the axio-dilaton away from its background value $\tau_s$, the first step is to recognize that $\cA_{D^2F^4}(\tau_s)$ and higher point terms involving $\delta \tau_s$ should all come from a single function 
\es{step1}{
g(\tau_s+\delta\tau_s,\bar\tau_s+\delta\bar\tau_s)=g(\tau_s,\bar\tau_s)+ \partial_{\tau_s}g(\tau_s,\bar\tau_s)\delta \tau_s + \partial_{\bar{\tau}_s}g(\tau_s,\bar\tau_s)\delta \bar{\tau}_s + \partial_{\tau_s} \partial_{\bar{\tau}_s} g(\tau_s,\bar\tau_s)\delta \tau_s \delta \bar{\tau}_s + \dots\,.
}
The next step is to realize that $\delta \tau_s \delta \bar{\tau}_s g(\tau_s,\bar\tau_s)$ cannot be promoted to a supervertex, because it is neither an F-term nor a D-term. The associated six-point function must therefore factorize into the lower-point functions which \textit{are} genuine supervertices upon taking soft limits. Since there are no candidates for the latter besides $D^2 F^4$ and $R$, this implies that
\es{diff}{
4\tau_2^2 \partial_{\tau_s} \partial_{\bar{\tau}_s} g(\tau_s, \bar{\tau}_s) = r(r - 1) g(\tau_s, \bar{\tau}_s)\,,
}
where $r = \frac{3}{2}$ is then fixed by comparing to the leading perturbative term $\tau_s^{3/2}$ in the Veneziano amplitude \eqref{V}. The solution to this equation is a non-holomorphic Eisenstein series 
 \es{EisensteinExpansion}{
  g(\tau_s) =-32\pi^5 E_{3/2}(\tau_s)
   &= -2^{6}\pi^5 \zeta(3)\tau_2^{3/2} -\frac{2^{6}\pi^7}{3\sqrt{\tau_2}} -2^{7}\pi^6\sqrt{\tau_2}\sum_{k\ne 0} \abs{k}
    \sigma_{-2}(|k|) \, 
      K_{1} (2 \pi \tau_2 \abs{k}) \, e^{2 \pi i k \tau_1} \, ,
 }
 where the divisor sum $\sigma_p(k)$ is defined as $\sigma_p(k)=\sum_{d>0,{d|k}}  d^p$, and $K_{1}$ is the Bessel function of the second kind. Note that this modular function does not factorize into holomorphic functions and has corrections at each genus order to the instantons, unlike the Jacobi theta functions for the $F^4$ term which are holomorphic and only have genus-zero contributions. We can then combine the known $\cA_{D^2F^4}(\tau_s)$ along with the localization constraints to completely fix the $M_{D^2F^4}(\tau_s)$ coefficients in \eqref{melGen2} to 
\es{D2F4np}{
\tilde{a}^2_{D^2F^4}(\tau_s)=-\frac{6\sqrt{2}}{\pi^{3/2}}E_{3/2}(\tau_s)\,,\qquad a^1_{D^2F^4}=\tilde a^1_{D^2F^4}=0\,.
}
The higher derivative corrections beyond $D^2F^4$ are unprotected, and we do not expect to be able to fix them from a protected quantity like the mass deformed sphere free energy.

%

\section{Conclusion}
\label{conc}

In this work, we computed the first few higher derivative corrections to gluon scattering on D7 branes using supersymmetric localization. In particular, we computed derivatives of the mass deformed sphere free energy $F(\mu_i)$ to all orders in $1/N$, and found that only the leading term depends on the complexified coupling $\tau$ via Jacobi theta functions. We then used the relation between these derivatives and the integrated holographic correlator to fix the first few $1/N$ corrections. At finite coupling, we fixed the $F^4$ correction in terms of Jacobi theta functions, which matched the known flat space limit, and then combined the flat limit with localization constraints to fix the $D^2F^4$ correction in terms of the non-holomorphic Eisenstein $E_{3/2}(\tau)$. At weak coupling, we found that the genus-zero holographic correlator is very similar to the flat space Veneziano amplitude, at least up to the $D^2F^4$ correction that we computed.

There are a couple of aspects of the calculation that deserve further study. While the AdS $F^4$ correction proportional to $1/N$ could be computed from localization and precisely matched to the flat space limit, some of the information needed for the next order is still missing. The 1-loop AdS gluon exchange term was computed in \cite{Alday:2021ajh} and further verified in \cite{Behan:2022uqr,Huang:2023oxf} but there is also a tree level graviton exchange term. The expression written in \cite{Alday:2021ajh} is missing the contributions from graviton KK modes that appear in long multiplets, and does not have the expected scaling in the flat space limit. In this paper we gave a prediction for this amplitude in the flat space limit and
we are confident that it is correct, since its $\log[\ell_s^2]$ threshold term contributes to the $F^4$ term at finite $\tau$, which we precisely matched. 
If this calculation can be completed, then we will be able to derive the $\tau$-independent contributions to flat space $F^4$ (which are the same order as 1-loop in the finite $\tau_s$ expansion) by fixing the corresponding AdS term using localization and taking the flat space limit. It would be nice to also compute these terms independently from type IIB string theory in the presence of D7 branes,\footnote{As discussed before, this was done when the transverse space is a torus, but not when it is flat, and the flat space limit of the torus is subtle in this case due to dependence on the complex modulus $U$ \cite{Kiritsis:2000zi}. This $U$ dependence factorizes from the $\tau_s$ dependence, though, so it is easy to isolate the $\tau_s$-dependent terms. } perhaps using the recent string field methods \cite{Agmon:2022vdj,Sen:2021tpp}.

As discussed in the main text, all the $\tau$ dependence at large $N$ comes from a novel $N$-independent contribution $Z_\text{extra}$ that was first discovered in 5d \cite{Kim:2012gu,Hwang:2014uwa}. This $Z_\text{extra}$ is also necessary in 4d, as can be checked by demanding that the $USp(2)$ theory be equivalent to $SU(2)$ SQCD, but has not been studied before. In our work, we conjectured a prescription for $Z_\text{extra}$ to 2-instanton order that passes various consistency checks at finite $N$, and also gave the expected large $N$ result. It would be nice to find an all orders expression for this 4d $Z_\text{extra}$, as has been done in the analogous 5d case. 

There are also many future directions at both strong and weak coupling. At weak coupling, it would be nice to compute the AdS Veneziano amplitude to higher orders in $1/\lambda$ (\textit{i.e.} small $\ell_s$), and see if it continues to resemble the flat space Veneziano amplitude. Perhaps one can constrain these higher order terms by relating to non-perturbative in $\lambda$ terms and imposing certain number theoretic properties on the coefficients, as was done for AdS graviton scattering in $\mathcal{N}=4$ SYM \cite{Alday:2022uxp,Alday:2022xwz}. The AdS gluon scattering to the order we computed already seems much simpler than AdS graviton scattering, even though we only have half as much supersymmetry, so it is possible we might be able to guess the full AdS Veneziano amplitude at finite $\lambda$. There may also be relations between the AdS gluon and graviton scattering, as has been recently explored in flat space \cite{Chi:2021mio,Chen:2022shl,Chen:2023dcx}. 

At finite string coupling, it would be nice to find an explicit (\textit{i.e.} no integrals) expression for the quartic mass derivatives of $F(\mu_i)$ at finite $N$ and $\tau$, as was found for quadratic mass derivatives for $\mathcal{N}=4$ SYM \cite{Dorigoni:2021guq}. The all orders in $1/N$ expansion in our theory is actually much simpler than the $\mathcal{N}=4$ SYM case, as we only have $\tau$ dependence at leading order, and the instantons in the Jacobi theta functions have no higher genus corrections, unlike the modular functions that appeared for $\mathcal{N}=4$ SYM \cite{Chester:2019jas,Chester:2020vyz,Alday:2021vfb,Dorigoni:2022zcr}. This suggests that a finite $N,\tau$ expression might be easier for quartic derivatives in our case. This quantity would be useful for numerically bootstrapping our theory at finite $N$, generalizing the $N=1$ case considered in \cite{Chester:2022sqb}. The numerical bootstrap will be essential for studying higher derivative corrections beyond the protected $F^4$ and $D^2F^4$ terms considered in this paper.

The results of this work can also be extended to other holographic 4d $\mathcal{N}=2$ theories. In this paper we considered the simplest F-theory construction with a holographic CFT dual, which is the unique case where $\tau_s$ can take any value,\footnote{There are a handful of others if one considers 4d $\mathcal{N} = 2$ S-folds \cite{Apruzzi:2020xxx} whose correlators are the same at tree level but different at 1-loop \cite{Behan:2022uqr}.} and in particular has a weak coupling regime. The other six known cases freeze $\tau_s$ to a strongly coupled value \cite{Dasgupta:1996ij}, and correspond to CFTs with the flavor symmetry groups \cite{Fayyazuddin:1998fb}
\es{Gs}{
G_F=\text{none}\,, SU(2)\,,SU(3)\,,E_6\,,E_7\,,E_8\,. 
}
These theories are all non-Lagrangian, with the first three cases being Argyres-Douglas type CFTs \cite{Argyres:1995jj,Argyres:1995xn}, and the last three being Minihan-Nemeschansky type CFTs \cite{Minahan:1996fg,Minahan:1996cj}. While localization can only be applied to $\mathcal{N}=2$ theories with Lagrangians, it may still be possible to study these theories using localization by considering them as infinite points on the Coulomb branch of Lagrangian theories \cite{Bissi:2021rei,Russo:2014nka,Russo:2019ipg}. This would give us the first window on intrinsically strongly coupled F-theory. 

Lastly, one can consider the constraints of localization on holographic correlators of other protected multiplets with half-maximal supersymmetry in various dimensions. For instance, the stress tensor correlator is dual to graviton scattering, and can be constrained by taking derivatives of the squashed sphere free energy, which has been computed using localization for certain half-maximal supersymmetric 3d CFTs \cite{Chester:2021gdw,Chester:2020jay}. One could also consider correlators of half-BPS Coulomb branch operators in 4d $\mathcal{N}=2$ theories, which are dual to certain KK modes.\footnote{For instance, it is straightforward to compute the OPE coefficient defined in \cite{Beem:2014zpa}: \es{coulombOPE}{|\lambda_{\mathcal{E}_2 \mathcal{E}_2 \mathcal{E}_4}|^2 = 2 + \frac{\partial_\tau \partial_{\bar{\tau}} \log \left ( \partial_\tau \partial_{\bar{\tau}} \log Z \right )}{\partial_\tau \partial_{\bar{\tau}} \log Z} = 2 + \frac{32}{(4N + 1)(4N + 3)}+O(e^{-N})\,,} which is remarkably independent of $\tau$ to all orders in $1/N$ (but won't be at finite $N$).} A certain protected OPE coefficient in these correlators can be computed from localization using the so-called $tt^*$ equations \cite{Baggio:2014sna,Baggio:2015vxa,Baggio:2014ioa,Gerchkovitz:2016gxx,Grassi:2019txd}, which has been explicitly computed at large $N$ for a variety of theories \cite{Billo:2022gmq,Billo:2022fnb,Billo:2022lrv}. This OPE coefficient could be imposed on the large $N$ correlator to fix some higher derivative corrections.

\section*{Acknowledgments} 

We thank Ofer Aharony, Xi Yin, Simon Caron-Huot, Silviu Pufu, Congkao Wen, Ying-Hsuan Lin, Xinan Zhou, Ellis Yuan, Fernando Alday, Satoshi Nawata, Zohar Komargodski, Francesco Fucito, and Jaewon Song for useful conversations, and Ofer Aharony, Xi Yin, and Fernando Alday for reviewing the manuscript. We are grateful to the organizers of the Bootstrap 2022 workshop in Porto where this project was conceived and acknowledge valuable collaboration which took place at the KITP program Bootstrapping Quantum Gravity. CB received funding from the European Research Council (ERC) under the European Union's Horizon 2020 research and innovation programme (grant agreement No 787185). SMC is supported by the Center for Mathematical Sciences and Applications and the Center for the Fundamental Laws of Nature at Harvard University. This research was supported in part by the National Science Foundation under Grant No. NSF PHY-1748958.

\appendix
\section{Matrix model details}
\label{tseytlin}

The basic ingredients and results of the matrix model integrals arising from supersymmetric localization were introduced in Sections \ref{perLoc} and \ref{perLoc2}. In this appendix we give more details on the matrix model computations, first in section \ref{pert} for large $N$ and large 't Hooft coupling $\lambda_\text{UV}\equiv g_\text{YM}^2N$, where the instanton term can be neglected; then for large $N$ but finite coupling $g_\text{YM}$ in section \ref{inst}, where instantons become relevant.

\subsection{Perturbative results}\label{pert}

Let us begin here with the perturbative terms: we consider $N\to \infty$ and $\lambda_\text{UV}\to \infty$, neglecting exponentially suppressed terms. We thus neglect the contribution of instantons in the partition function \eqref{Znew}.
Recalling \eqref{Zloop}, this makes the $S^4$ partition function an expectation of $e^{-S_{\text{int}}(X, \mu_i)}$ with respect to a Gaussian matrix model. First, note that we are only interested in terms up to $O(\mu^4)$, since ultimately we want to compute four mass derivatives of the free energy and set the masses to zero. So we can expand
\begin{align}
S_{\text{int}}(X,\mu_i)=S^{(0)}_{\mathrm{int}}+\sum_{k=1}^{\infty}m_{(2k)}\,S^{(k)}_{\mathrm{int}}\,,
\end{align}
where we have introduced
\begin{align}
m_{(2k)}=\frac{1}{4}\sum_{i=1}^4\mu_i^{2k}\,,
\end{align}
and the relevant terms for our computations are
\begin{align}
S^{(0)}_{\mathrm{int}}=-2\sum_{n=1}^N\log\frac{H(2x_n)}{H(x_n)^4}\,,\quad 
S^{(1)}_{\mathrm{int}}=4\sum_{n=1}^N\partial^2_{x_n}\log H(x_n)\,,\quad
S^{(2)}_{\mathrm{int}}=\frac{1}{3}\sum_{n=1}^N\partial^4_{x_n}\log H(x_n)\,.
\end{align}
Following \cite{Beccaria:2022kxy}, we define the $USp(2N)$ $\mathcal{N}=4$ free energy
\begin{align}
\begin{split}
e^{-F_{\mathcal{N}=4}}&=\int [dX]\,e^{-\frac{8\pi^2}{g_\text{YM}^2}\,\text{tr}X^2}\,,\\
F_{\mathcal{N}=4}&=\frac{1}{2}N(2N+1)\log\left(\frac{16\pi^2N}{\lambda_{\text{UV}}}\right)+\log\frac{G(3/2)}{G(1+N)G(3/2+N)}\,,
\end{split}
\end{align}
and write the $\mathcal{N} = 2$ free energy as
\begin{align}\label{FN2}
F(\mu_i)=F_{\mathcal{N}=4}+\Delta F(\mu_i)\,,\qquad \Delta F=N\,F_1+F_2+\frac{1}{N}F_3+\ldots\,,
\end{align}
where each term has the expansion
\begin{align}
F_j = \sum_{k = 1}^\infty m_{(2k)}F_j^{(k)}.
\end{align}
The brute force way to compute the free energy is then to use the cumulant expansion
\es{cumulant}{
\Delta F = \langle S_{\mathrm{int}} \rangle - \frac{1}{2!} \langle S_{\mathrm{int}}^2 \rangle_\text{con} + \frac{1}{3!} \langle S_{\mathrm{int}}^3 \rangle_\text{con} - \dots
}
up to the number of terms desired in \eqref{FN2}. This requires $1/N$ expansions for the connected multi-point correlators of $\text{tr} X^{2m}$ which may be found in \cite{Beccaria:2021ism}. Although we will discuss a more efficient method momentarily, one basic structural property of this result which should be emphasized is that the two-point correlator is special. Sums over $m_i$ of $\langle \text{tr} X^{2m_1} \text{tr} X^{2m_2} \text{tr} X^{2m_3} \rangle_\text{con}$ and higher factorize whereas for the leading $\langle \text{tr} X^{m_1} \text{tr} X^{m_2} \rangle_\text{con}$, which affects $F_2$, one needs to use an integral representation to deal with a factor of $1/(m_1 + m_2)$.\footnote{This can be seen from the prefactor in the expression
\es{multipoint}{
\langle \text{tr} X^{2m_1} \dots \text{tr} X^{2m_M} \rangle_\text{con} = \frac{2 (\sum_{i = 1}^M m_i - M)_M}{N^{M - 2}} \prod_{i = 1}^M \left ( \frac{\lambda_\text{UV}}{4\pi^2 N} \right )^{m_i} \frac{\Gamma(m_i + \tfrac{1}{2})}{\sqrt{\pi} \Gamma(m_i)} \left [ 1 + \sum_{j = 1}^\infty \frac{p_j(m_i)}{N^j} \right ],
}
with $p_j(m_i)$ being polynomials.
}

For this matrix model, there is fortunately a fact which makes our lives easier.
Because all interaction terms are single-trace deformations of a Gaussian matrix model, the higher order terms can be computed recursively from lower order ones using the so-called Toda equations \cite{Beccaria:2022kxy}. In terms of the variable $y \equiv (4\pi)^2 N / \lambda_\text{UV}$, these read
\es{toda}{
\log \left ( -\partial_y^2 F \right ) = 2F - F\bigl|_{N \to N + 1} - F\bigl|_{N \to N - 1}
}
where the shift in $N$ occurs after $\lambda_\text{UV}$ is substituted for its expression in terms of $y$. Following \cite{Beccaria:2022kxy} who defined
\es{todaSol1}{
\mathcal{F} = \frac{\partial^2}{\partial \lambda_\text{UV}^2} (\lambda_\text{UV} F_1),
}
we can expand both sides of \eqref{toda} in $1/N$ to get a series of equations
\es{todaSol2}{
F_2^\prime &= \frac{1}{4} \mathcal{F} (1 - \lambda_\text{UV} \mathcal{F}), \\
F_3 &= \frac{\lambda_\text{UV}^2}{48} (-3\mathcal{F}^2 + 2\lambda_\text{UV} \mathcal{F}^3 + \mathcal{F}^\prime), \dots
}
which show that almost all $F_j$ are determined in terms of $F_1$. The exception is $F_2$ since one must compute it using an integral which is similar to the complication caused by the two-point correlator in the brute force approach. The constant of integration should be fixed by a weak coupling comparison which involves going to second order in \eqref{cumulant} and extracting only the leading power of $\lambda_\text{UV}$ which does not require an infinite sum.

The first step is thus to solve for
\begin{align}
F_1=\lim_{N\to \infty} \frac{1}{N} \langle S_{\mathrm{int}}\rangle.
\end{align}
We provide an alternative route to the computation of this expectation value, following {\it e.g.} \cite{Binder:2019jwn}. Consider for instance 
\begin{align}
\langle S^{(0)}_{\mathrm{int}}\rangle=-2\sum_{n=1}^N \langle\log H(2x_n)\rangle+8 \sum_{n=1}^N\langle\log H(x_n)\rangle\,.
\end{align}
One can use \cite{Russo:2013kea}
\begin{align}
\log H(x)=\int_0^{\infty}d\omega\frac{1-2x^2\omega^2-\cos(2\omega x)}{2\omega \sinh^2\omega}\,,
\end{align}
to rewrite the expectation value as
\begin{align}
\langle S^{(0)}_{\mathrm{int}}\rangle=\int_0^{\infty}d\omega\frac{8}{\omega\sinh^2\omega}\sum_{n=1}^N\langle\sin^4(\omega x_n)\rangle\,.
\end{align}
To compute the large $N$ limit of expectation values in a Gaussian matrix model, it is convenient to use Wigner's semi-circle law: in the large $N$ limit, the distribution of the eigenvalues of $X$ converges to a semi-circular distribution and the leading order contribution to the expectation values of a function $f(x_n)$ of the eigenvalues is given by
\begin{align}
\sum_{n=1}^N \langle f(x_n)\rangle\approx \frac{2N}{\pi}\int_{-1}^1dx\,\sqrt{1-x^2}\,f\left(\frac{\sqrt{\lambda_{\text{UV}}}}{2\pi}x\right)\,.
\end{align}
In the case of interest, we find
\begin{align}
\langle S^{(0)}_{\mathrm{int}}\rangle = \int_0^{\infty}d\omega\frac{8}{\omega\sinh^2\omega}\,\left[\frac{2N}{\pi}\int_{-1}^1dx \sqrt{1-x^2}\sin^4\left(\frac{\sqrt{\lambda_{\text{UV}}}}{2\pi}\omega\,x\right)\right]+O(1)\,.
\end{align}
Then, using
\begin{align}
\frac{2}{\pi}\int_{-1}^1\sqrt{1-x^2}\,e^{iax}=\frac{2}{a}J_1(a)\,,
\end{align}
we obtain
\begin{align}\label{F1nomass}
F_1^{(0)}=\lim_{N\to \infty}\frac{1}{N}\langle S^{(0)}_{\mathrm{int}}\rangle=\frac{\pi}{\sqrt{\lambda_\text{UV}}}\int_0^{\infty}\frac{d\omega}{\omega^2\sinh^2\omega}\Big[\frac{3\sqrt{\lambda_\text{UV}}\omega}{\pi}-8J_1\big(\frac{\sqrt{\lambda_\text{UV}}\omega}{\pi}\big)+J_1\big(\frac{2\sqrt{\lambda_\text{UV}}\omega}{\pi}\big)\Big]\,.
\end{align}

This result was also found in \cite{Beccaria:2022kxy} by using the weak coupling expansion where the integral comes from
\es{zetaInt}{
\zeta(2n + 1) = \frac{1}{4\Gamma(2n + 2)} \int_0^\infty \frac{d\omega \, \omega^{2n + 1}}{\sinh^2(\omega / 2)}.
}
Since the expression of $S^{(k)}_{\mathrm{int}}$ simply involves derivatives of $\log H(x)$, either method can be used to find
\begin{align}\label{F1mass}
\begin{split}
F_1^{(1)}&=\frac{8\pi}{\sqrt{\lambda_{\text{UV}}}}\int_0^{\infty}\frac{d\omega}{\sinh^2\omega}\Big[2J_1\big(\frac{\sqrt{\lambda_{\text{UV}}}\omega}{\pi}\big)-\frac{\sqrt{\lambda_{\text{UV}}}\omega}{\pi}\Big]\,,\\
F_1^{(2)}&=-\frac{16\pi}{3\sqrt{\lambda_{\text{UV}}}}\int_0^{\infty}\frac{d\omega}{\sinh^2\omega}\omega^2 J_1\big(\frac{\sqrt{\lambda_\text{UV}}\omega}{\pi}\big)\,,
\end{split}
\end{align}
as well. The utility of these results is that one can insert the Mellin representation of the Bessel function
\es{Bessel}{
J_1(z) = \int \frac{ds}{2\pi i} \frac{\Gamma(-s)}{\Gamma(s + 2)} \left ( \frac{z}{2} \right )^{2s + 1}\,,
}
and close the contour to the left to generate a strong coupling expansion with only finitely many terms.\footnote{This property would be obscure if one used the zeta function integral representations based on $\sinh^{-1}(t/2)$ or $\cosh^{-1}(t/2)$. It is manifest with \eqref{zetaInt} because $\zeta(2n + 1)\Gamma(2n + 2)$ is regular when $n$ is a negative half-integer.}
For identical $\mu_i$, \cite{Beccaria:2022kxy} applied the Toda equation to \eqref{F1nomass} and \eqref{F1mass} to derive subleading orders in the large $N$ expansion and guess an expression for the free energy that resums all orders in the large $N$ and large $\lambda_\text{UV}$, up to exponentially small contributions. After correcting an error in that expression for the $N^0$ term, we then generalize it by just replacing powers of their mass parameter $m$ with a suitable combination of our $m_{(2k)}$. The result is
\es{slightTseyt}{
F_{\text{pert}}&= \left(N+\frac{3}{4}+2 m_{(2)}\right)\left(N+\frac{1}{4}+2 m_{(2)}\right) \log \frac{16 \pi^2 N}{\lambda} \\
& -\log \left[G\left(N+\frac{5}{4}+2 m_{(2)}\right) G\left(N+\frac{7}{4}+2 m_{(2)}\right)\right]-8m_{(2)}^2\,\log(4\pi)
-\frac13m_{(4)} \Big[\frac{16 \pi^2 N}{\lambda}-8\log2\Big]\\
&+   (8 \zeta (3)-\frac{44}{3}-8 \gamma +8 \log (4 \pi ))m_{(2)}^2+O(\mu^6)
\,,
}
where $\lambda$ is related to $\lambda_\text{UV}$ by the perturbative part of the UV/IR relation \eqref{AdSV}
\begin{align}
\frac{1}{\lambda}=\frac{1}{\lambda_\text{UV}}+\frac{\log 2}{2\pi^2N}\,,
\end{align}
and we dropped physically meaningless $\lambda$-independent $\mu^0$ and $\mu^2$ terms.\footnote{These terms are ambiguous due to the Weyl anomaly.} From the above, it is straightforward to obtain the result \eqref{largeLamF}. 

The $m_{(2)}^2$ term on the last line of \eqref{slightTseyt} is the $N^0$ terms that \cite{Beccaria:2022kxy} missed, suitably generalized to different masses. This term is subtle, since it would be missed if one first expands $F_1$ at large $\lambda$, and then uses the Toda equation \eqref{todaSol2} to compute $F_2$ by taking the integral in $\lambda$. Instead, one must take large $\lambda$ only after computing $F_2$ at finite $\lambda$. We demonstrate this for the $m_{(2)}^2$ term $F_2^{(2,2)}$ at subleading order in $1/N$, which can be computed as
\es{F2m2}{
F_2^{(2,2)}&=-\frac14\int_0^{\lambda_\text{UV}} d\lambda_\text{UV}\, \lambda_\text{UV}((\lambda_\text{UV} F_1^{(2)})'')^2\\
&=\int_0^\infty dwd\omega\frac{8 \sqrt{\lambda_\text{UV} } w^2 \omega^2  \left(w J_0\left(\frac{w
   \sqrt{\lambda_\text{UV} }}{\pi }\right) J_1\left(\frac{\omega
   \sqrt{\lambda_\text{UV} }}{\pi }\right)-\omega J_1\left(\frac{w
   \sqrt{\lambda_\text{UV} }}{\pi }\right) J_0\left(\frac{\omega
   \sqrt{\lambda_\text{UV} }}{\pi }\right)\right)}{\pi \sinh^2w\sinh^2\omega
   \left(w^2-\omega^2\right)}\,,\\
}
where we used the expression for $F_1^{(2)}$ in \eqref{F1mass}. We see that the $\lambda_\text{UV}$ integral made the resulting $w,\omega$ dependence not factorize, unlike all other $F_n$ for $n>1$. So for $n=2$ we cannot evaluate large $\lambda_\text{UV}$ by simply applying \eqref{Bessel}, then taking all auxiliary integrals and evaluating poles, which works for $n\neq2$. Instead, we adapt a strategy used in a similar context in \cite{future}. We start by using the Bessel kernel identity to rewrite the integral as
\es{BesKern}{
F_2^{(2,2)}&=-\sum_{k=1}^\infty \int_0^\infty dwd\omega\frac{32 k w \omega}{\sinh^2w\sinh^2\omega}
   J_{2 k}\left(\frac{w \sqrt{\lambda_\text{UV} }}{\pi }\right) J_{2
   k}\left(\frac{\omega \sqrt{\lambda_\text{UV} }}{\pi }\right)\\
   &=-\sum_{k=1}^\infty\int \frac{ds dt}{(2\pi i)^2}\frac{k 
   \zeta (2 s+3) \Gamma (2 s+4) \zeta (2t+3) \Gamma
   (2t+4) \Gamma (k-s-1) \Gamma (k-t-1)
   \lambda_\text{UV} ^{s+t+2}}{  2^{4 s+4t+3} \pi ^{2 (s+t+2)} \Gamma (k+s+2) \Gamma
   (k+t+2)}\,,\\
}
where in the second line we used \eqref{Bessel} and \eqref{zetaInt}, and we shifted the contours by $s\to s-k+1$ and $t\to t-k+1$. The contour integral includes contributions from two kinds of poles: separate poles in $s,t$, and simultaneous poles in $s+t$. We denote the first as $I$, which we get by performing the $k$ sum in \eqref{BesKern} to get
\es{I}{
I&=-\int \frac{ds dt}{(2\pi i)^2}\frac{ \zeta (2 s+3) \Gamma
   (-s) \Gamma \left(s+\frac{5}{2}\right) \zeta (2 t+3)
   \Gamma (-t) \Gamma \left(t+\frac{5}{2}\right) \lambda_\text{UV}
   ^{s+t+2}}{2^{2 (s+t-1)} \pi ^{2 s+2 t+5}(s+t+2)}\\
   &=-4 (\log \lambda_\text{UV} +2 \gamma +2-2 \log (4 \pi ))\,,
}
where in the second line we took the pole $s=t=-1$. The second kind of contribution $II$ from simultaneous $s+t$ poles can be obtained by expanding \eqref{BesKern} at large $k$ and doing the sum order by order to get
\es{II}{
II&=-\int \frac{ds dt}{(2\pi i)^2}\frac{ \zeta (2 s+3) \Gamma (2
   (s+2)) \zeta (2 t+3) \Gamma (2 (t+2)) \lambda_\text{UV} ^{s+t+2}
   \zeta (2 s+2 t+5)}{2^{4 s+4 t+3} \pi ^{2 (s+t+2)}}+\dots\\
   &=-\int \frac{dt}{2\pi i}16 \zeta (-2 t-1) \zeta (2 t+3) \Gamma (-2 t) \Gamma (2
   (t+2))\,,
}
where in the second line we took the pole $s+t=-2$, and we could drop the subleading large $k$ sums given by the ellipses because they have no poles at negative $s+t$. If we naively now take the poles with $t=-n$ for $n=1,2,\dots$ we would get
\es{II2}{
\widetilde{II}=-\sum_{n=1}^\infty \frac{8 \zeta (3-2 n) \zeta (2 n-1) \Gamma (2 n)}{\Gamma (2
   n-3)}\,.
}
This sum is divergent, unlike the integral in \eqref{II} which we can numerically check converges. But we can regulate the divergence in \eqref{II2} using the identities
\es{idents}{
\zeta(x)=\frac{1}{\Gamma(x)}\int_0^\infty \frac{y^{x-1}}{e^{y}-1}dy\,,\qquad \zeta(3-2n)=-2\frac{\Gamma(2n-2)\zeta(2n-2)\cos(\pi n)}{(2\pi)^{2n-2}}\,,
}
and then doing the sum over $n$ to get
\es{II3}{
II&=\int_0^\infty dxdy\Big[\frac{12 x y^2 \cos \left(\frac{x y}{2 \pi }\right)}{\pi ^2
   \left(e^x-1\right) \left(e^y-1\right)}-\frac{2 x^2 y^3
   \sin \left(\frac{x y}{2 \pi }\right)}{\pi ^3
   \left(e^x-1\right) \left(e^y-1\right)}\Big]\\
   &=-\frac{20}{3} + 8 \zeta(3)\,,
}
where we can check numerically that this equals the integral in \eqref{II}, which justifies our regularization. Adding \eqref{II} to \eqref{I} then gives the last line of \eqref{slightTseyt}, where the $\log\lambda_\text{UV}$ term was already included in the previous lines. Note that higher orders in mass will also receive contributions at order $F_2$ of the same type, even if all $F_n$ at large $\lambda$ for $n>2$ are zero, but we only work to $O(m^4)$.

\subsection{Instantons}\label{inst}

So far our results are just a trivial generalization of \cite{Beccaria:2021ism,Beccaria:2022kxy}. The main novelty concerns computations at finite $g_{\text{YM}}$, which require taking into account the contribution of instantons. In this section, we will define $q_{\text{UV}}=e^{2\pi i \tau_{\text{UV}}}$ and $q=e^{2\pi i \tau}$, related to each other by \eqref{UVtoIR}, and write
\es{zinstSum}{
Z_{\text{inst}}(X,\mu_i,\tau_{\text{UV}})=\sum_{k=0}^{\infty}q_{\text{UV}}^k\,Z_k(X,\mu_i)\,,
}
while stressing again that this is not yet the full instanton partition function. The $k$-instanton contribution to $Z_{\text{inst}}$ is built out of contributions from all supermultiplets present in the theory. The individual contributions from the multiplets of interest can be obtained from \cite{Shadchin:2005mx} where they are expressed in terms of zero modes $\phi_I$ for the ADHM theory which appears in the 5d construction. They also depend on the antisymmetric hypermultiplet mass $m$ and general $\Omega$-deformation parameters $\epsilon_{1,2}$. At the end of the day we will set $m = 0$ and $\epsilon_{1,2} = 1$ but it is useful to keep such parameters generic in intermediate steps of the computations.
Adapted to our conventions\footnote{In particular we set
\begin{align}
\mu_i^{\text{there}}=\mu_i^{\text{here}}+\frac{\epsilon_1+\epsilon_2}{2}\,,\quad 
m^{\text{there}}=m^{\text{here}}+\frac{\epsilon_1+\epsilon_2}{2}\,, \quad
a^{\text{there}}=i x^{\text{here}}\,.
\end{align}}
the building blocks of \cite{Shadchin:2005mx} are
\begin{align}\label{z_kinst}
\begin{split}
z^{\text{hyper($i$)}}_k=&\mu_i^\chi\,\prod_{I=1}^n(\mu_i^2 - \phi_I^2)\,,\\
z^{\text{vec}}_k=&\frac{(-1)^n}{2^{n+\chi}n!}\frac{\epsilon_+^n}{\epsilon_1^n\epsilon_2^n}\,\left[\frac{1}{\epsilon_1\epsilon_2\prod_{l=1}^N(\epsilon_+^2/4+x_l^2)}\prod_{I=1}^n\frac{\phi_I^2(\phi_I^2-\epsilon_+^2)}{(\phi_I^2-\epsilon_1^2)(\phi_I^2-\epsilon_2^2)}\right]^{\chi}\\
&\times\left[\prod_{I<J\leq n}\frac{(\phi^-_{IJ})^2\,((\phi^-_{IJ})^2-\epsilon_+^2)\,(\phi^+_{IJ})^2\,((\phi^+_{IJ})^2-\epsilon_+^2)}{((\phi^-_{IJ})^2-\epsilon_1^2)\,((\phi^-_{IJ})^2-\epsilon_2^2)\,((\phi^+_{IJ})^2-\epsilon_1^2)\,((\phi^+_{IJ})^2-\epsilon_2^2)}\right]\\
&\times\prod_{I=1}^n\frac{1}{(4\phi_I^2-\epsilon_1^2)(4\phi_I^2-\epsilon_2^2)\prod_{l=1}^N((\phi_I-\epsilon_+/2)^2+x_l^2)((\phi_I+\epsilon_+/2)^2+x_l^2)}\,,\\
z^{\text{anti}}_k=&\frac{(m^2-\epsilon_-^2/4)^n}{(m^2-\epsilon_+^2/4)^n}\,\left[\frac{\prod_{l=1}^N(m^2+x_l^2)}{m^2-\epsilon_+^2/4}\prod_{I=1}^n\frac{(\phi_I^2-(m+\epsilon_-/2)^2)(\phi_I^2-(m-\epsilon_-/2)^2)}{(\phi_I^2-(m+\epsilon_+/2)^2)(\phi_I^2-(m-\epsilon_+/2)^2)}\right]^{\chi}\\
&\times\prod_{I<J\leq n}\frac{((\phi^-_{IJ})^2-(m+\epsilon_-/2)^2)((\phi^-_{IJ})^2-(m-\epsilon_-/2)^2)}{((\phi^-_{IJ})^2-(m+\epsilon_+/2)^2)((\phi^-_{IJ})^2-(m-\epsilon_+/2)^2)}\\
&\times\prod_{I<J\leq n}\frac{((\phi^+_{IJ})^2-(m+\epsilon_-/2)^2)((\phi^+_{IJ})^2-(m-\epsilon_-/2)^2)}{((\phi^+_{IJ})^2-(m+\epsilon_+/2)^2)((\phi^+_{IJ})^2-(m-\epsilon_+/2)^2)}\\
&\times \prod_{I=1}^n\frac{\prod_{l=1}^N((\phi_I-m)^2+x_l^2)((\phi_I+m)^2+x_l^2)}{(4\phi_I^2-(m+\epsilon_+/2)^2)(4\phi_I^2-(m-\epsilon_+/2)^2)}\,,
\end{split}
\end{align}
where we have introduced the combinations
\begin{align}
\epsilon_{\pm}=\epsilon_1\pm\epsilon_2\,,\quad \phi^{\pm}_{IJ}=\phi_I\pm\phi_J\,,
\end{align}
and defined $n \equiv \lfloor k/2 \rfloor$ and $\chi \equiv k \; \text{mod} \; 2$ so that $k = 2n + \chi$.
These expressions allow the $k$-instanton terms in \eqref{zinstSum} to be computed via the contour integrals \eqref{Zdetail} which we restate as
\es{JKintegral}{
Z_k(X,\mu_i)=\int \prod_{I=1}^n\frac{d\phi_I}{2\pi i}z_k^{\text{anti}}(X,\phi)z_k^{\text{vec}}(X,\phi)\prod_{i=1}^{N_f}z_k^{\text{hyper}(i)}(X,\phi,\mu_i)\,.
}
The so-called Jeffrey-Kirwan contour \cite{jeffrey1994localization} which must be used was discussed for 4d theories without antisymmetric hypermultiplets in \cite{Hollands:2010xa} and later explained more generally in 5d \cite{Hwang:2014uwa}. It involves summing over residues that have $\epsilon_{1,2}$ appearing wth a fixed sign which should be positive for $z_k^{\text{vec}}$ if and only if it is negative for $z_k^{\text{anti}}$. The reduction to 4d is trivial and simply consists in discarding poles with negative $e^{\phi_I}$.

As reviewed in \cite{Avraham:2018xxx}, the ADHM gauged quantum mechanics from which \eqref{z_kinst} were derived can contain extra degrees of freedom which should be removed to obtain the correct instanton partition function \cite{Kim:2012gu,Hwang:2014uwa}. For the antisymmetric hypermultiplet, these come from the singlet which forms the full representation for $N = 1$ and an irreducible component of it for $N > 1$. They should be removed
by a partition function $Z_{\text{extra}}$, which does not depend on the $USp(2N)$ eigenvalues, such that the actual partition function of the theory is obtained dividing the naive expression by $Z_{\text{extra}}$ and expressing the result in terms of the IR coupling $\tau$, as described by \eqref{Znew}. The origin and explicit expression of $Z_{\text{extra}}$ was understood for 5d $\mathcal{N}=1$ SCFTs with $USp(2N)$ gauge group, 1 antisymmetric hypermultiplet and $0\leq N_f\leq 7$ fundamental hypermultiplets, where the enhancement of the global symmetry to $E_{N_f+1}$ is only observed in the partition function after the contribution of $Z_{\text{extra}}$ is properly accounted for \cite{Kim:2012gu,Hwang:2014uwa,Chen:2023smd}. However, to the best of our knowledge the 4d version of this story has been far less explored and explicit results for $Z_{\text{extra}}$ are not available in the literature.\footnote{A subtlety with trying to naively dimensionally reduce $Z^{\text{5d}}_\text{extra}$ to 4d is visible in
\es{5dextra}{
Z^{\text{5d}}_\text{extra} \approx 1 - \frac{q_\text{UV} + \bar{q}_\text{UV}}{2 \sinh \left ( \frac{\epsilon_1}{2} \right ) \sinh \left ( \frac{\epsilon_2}{2} \right ) \sinh \left ( \frac{\epsilon_+ \pm 2m}{4} \right )} \left [ \prod_{i = 1}^{N_f} \sinh \left ( \frac{\mu_i}{2} \right ) + \prod_{i = 1}^{N_f} \cosh \left ( \frac{\mu_i}{2} \right ) \right ].
}
When $N = 1$, the factors of $\sinh \left ( \frac{\mu_i}{2} \right )$ fully cancel the antisymmetric hypermultiplet terms which dimensionally reduce to the contributions of \eqref{z_kinst}. So discarding the factors of $\cosh \left ( \frac{\mu_i}{2} \right )$ is a valid prescription at 1-instanton but not at 2-instantons when there is no longer a clear distinction between these terms.
}

Despite the lack of an explicit expression for $Z_{\text{extra}}$, various arguments directly in 4d for why it must exist have been discussed in the main text. To recap, a naive application of \eqref{JKintegral} for $N = 1$ does not agree with the known $SU(2)$ SQCD partition function \cite{Alday:2009aq} or recover the UV/IR relation
\es{UVtoIR2}{
\pi i \tau = - \frac{\pi K(1 - q_\text{UV}^2 / 16)}{2 K(q_\text{UV}^2 / 16)} = 2\pi i \tau_\text{UV} - 4\log 2 + \frac{1}{64} q_\text{UV}^2 + \frac{13}{32768} q_\text{UV}^4 + O(q_\text{UV}^6)\,,
}
through \eqref{prepot} unless the antisymmetric hypermultiplet is dropped by hand. Here, $K(x)$ is the elliptic integral of the first kind. Additionally, numerical evaluations of \eqref{Z} both for $N = 1, 2$ do not lead to mass derivatives of the free energy 
which are exchanged by $SL(2, \mathbb{Z})$. Our prescription below for removing the $USp(2N)$ singlet will offer a solution to these problems, the latter of which has already been seen in Figure \ref{usp4check}. This makes it very different from the $U(1)$ factor of \cite{Alday:2009aq} which does not affect derivatives of the mass deformed sphere free energy. The factor relating $SU(2)$ to $USp(2)$ in \cite{Hollands:2010xa} drops out of mass derivatives as well (as it must for $SU(2) \cong USp(2)$ to hold) but nevertheless displays a superficial similarity to our factor in that its eigenvalue independence is only manifest when it is expressed in terms of the IR coupling.\footnote{Strictly speaking, the authors of \cite{Hollands:2010xa} only give this factor for the unrefined limit of $\epsilon_1 = -\epsilon_2$. For $\epsilon_1 = \epsilon_2 = 1$ of relevance here, we have found that their (2.24) should have $M = (\mu_1 + \mu_2 + 3)(\mu_3 + \mu_4 - 1) - \mu_1\mu_2 - \mu_3\mu_4 - \sum_i \mu_i - 3$ and $N = \frac{1}{2} \sum_i \mu_i^2 - \frac{11}{8}$.}

To obtain a 4d $Z_{\text{extra}}$ we then adopt the strategy outlined in the main text. Firstly, we assume that $Z_{\text{extra}}$ is $N$-independent as in 5d, so we can focus on $USp(2)$. Secondly, we demand that the partition function for $USp(2)$ be related to $SU(2)$ SQCD as shown in \cite{Hollands:2010xa}, such that $Z_{\text{extra}}$ is defined to cancel the residual effect of the antisymmetric hypermultiplet. In particular, for the $USp(2)$ theory we define
\begin{align}
\begin{split}
Z_{\text{anti}}&=\sum_{k=0}^{\infty}q_{\text{UV}}^k \int \prod_{I=1}^n\frac{d\phi_I}{2\pi i}z_k^{\text{anti}}z_k^{\text{vec}}\prod_{i=1}^{N_f}z_k^{\text{hyper}(i)}\,,\\
Z_{\text{no-anti}}&=\sum_{k=0}^{\infty}q_{\text{UV}}^k \int \prod_{I=1}^n\frac{d\phi_I}{2\pi i}z_k^{\text{vec}}\prod_{i=1}^{N_f}z_k^{\text{hyper}(i)}\,,
\end{split}
\end{align}
where for now we let $N_f\leq4$, and then define
\begin{align}
Z_{\text{extra}}\equiv \frac{Z_{\text{anti}}}{Z_{\text{no-anti}}}\,.
\end{align}
 For $N_f=0,1,2$ this prescription gives the closed form expressions
\begin{align}
\begin{split}
& N_f=0:\quad Z_{\text{extra}}=e^{-\frac{2q_{\text{UV}}}{\epsilon_1\epsilon_2(\epsilon_1+\epsilon_2+2m)(\epsilon_1+\epsilon_2-2m)}}\,,\\
& N_f=1:\quad Z_{\text{extra}}=e^{-\frac{2q_{\text{UV}}\,\mu_1}{\epsilon_1\epsilon_2(\epsilon_1+\epsilon_2+2m)(\epsilon_1+\epsilon_2-2m)}}\,,\\
& N_f=2:\quad Z_{\text{extra}}=e^{-\frac{2q_{\text{UV}}\,\left(\mu_1\mu_2+\tfrac{q_{\text{UV}}}{64}\right)}{\epsilon_1\epsilon_2(\epsilon_1+\epsilon_2+2m)(\epsilon_1+\epsilon_2-2m)}}\,,
\end{split}
\end{align}
which we checked up to 5 instantons. For $N_f=3$ we were not able to find a general expression, but up to 2 instantons we found
\es{eq1}{
&N_f=3: \quad Z_{\text{extra}}=e^{-\frac{2q_{\text{UV}}\,\mu_1\mu_2\mu_3}{\epsilon_1\epsilon_2(\epsilon_1+\epsilon_2+2m)(\epsilon_1+\epsilon_2-2m)}}\\
&\times \left[1+\frac{36m^2-5\epsilon_1^2-14\epsilon_1\epsilon_2-5\epsilon_2^2-16(\mu_1^2+\mu_2^2+\mu_3^2)}{512\epsilon_1\epsilon_2(\epsilon_1+\epsilon_2+2m)(\epsilon_1+\epsilon_2-2m)}\,q_{\text{UV}}^2+O(q_{\text{UV}}^3)\right]\,.\\
}
Lastly, we consider our main interest: $N_f=4$. Since the theory is now a CFT, $q_\text{UV}$ gets renormalized and becomes $q$ as given in \eqref{UVtoIR2}. The classical contribution to the holomorphic part of the partition function at finite $\epsilon_1,\epsilon_2$ is $e^{-\frac{2\pi i\tau_\text{UV}\tr X^2}{\epsilon_1\epsilon_2} }$. If we want to write this classical piece in terms of the IR $\tau$, as done in the second line of \eqref{Znew} for $\epsilon_1=\epsilon_2=1$, then this generates a shift in the instanton terms according to \eqref{UVtoIR2} so that we now define
\begin{align}
N_f=4:\qquad\qquad Z_{\text{extra}}=\frac{Z_{\text{anti}}}{Z_{\text{no-anti}}}\left(1+\frac{\sum_{l=1}^Nx_l^2}{64\epsilon_1\epsilon_2}q^2_{\text{UV}}\right)+O(q^3_{\text{UV}})\,,
\end{align}
where $x_l$ are the eigenvalues for $USp(2N)$. With this definition for $N=1$ we find
\es{eq2}{
&N_f=4:\quad Z_{\text{extra}}=e^{-\frac{2q_{\text{UV}}\,\mu_1\mu_2\mu_3\mu_4}{\epsilon_1\epsilon_2(\epsilon_1+\epsilon_2+2m)(\epsilon_1+\epsilon_2-2m)}}\\
&\times \Big[1-\frac{q_{\text{UV}}^2}{32\epsilon_1\epsilon_2(\epsilon_1+\epsilon_2+2m)(\epsilon_1+\epsilon_2-2m)}\\
&\hspace{0.5cm}\Big(\sum_{1\leq i<j\leq 4}\mu_i^2\mu_j^2-\frac{36m^2-5\epsilon_1^2-14\epsilon_1\epsilon_2-5\epsilon_2^2}{16}\sum_{i=1}^4\mu_i^2\\
&\hspace{0.5cm}+\frac{144m^4-111(\epsilon_1^4+\epsilon_2^4)-440\epsilon_1\epsilon_2(\epsilon_1^2+\epsilon_2^2)+472m^2(\epsilon_1^2+\epsilon_2^2)+800m^2\epsilon_1\epsilon_2-642\epsilon_1^2\epsilon_2^2}{256}\Big)\\
&\hspace{0.5cm}+O(q_{\text{UV}}^3,\mu_i^6)\Big]\,,
}
which is now eigenvalue independent as for $N_f<4$, and so makes sense as a general $N$ formula.

We can now consider the large $N$ and finite $\tau$ expansion of the instanton contributions to the mass derivatives of the free energy. Let us assume for a moment that $Z_{\text{inst}}$ is exponentially suppressed in this limit, so that only $Z_{\text{extra}}$ contributes. Since $ Z_{\text{extra}}$ does not have eigenvalue dependence, we can simply take mass derivatives of the $m=0,\epsilon_1=\epsilon_2=1$ expression in \eqref{ZextraFinal} to get 
\begin{align}\label{Ffinal2}
\begin{split}
-\partial_{\mu_1}\partial_{\mu_2}\partial_{\mu_3}\partial_{\mu_4}\log|Z_\text{extra}|^2\big|_{\mu=0}&=\frac{1}{2}(q_{\text{UV}}+\bar{q}_{\text{UV}})+O(q_{\text{UV}}^3,\bar{q}_{\text{UV}}^3)\\
&=8(q^{1/2}+\bar{q}^{1/2})+O(q^{3/2},\bar{q}^{3/2})\,,\\
-\partial_{\mu_1}^2\partial_{\mu_2}^2\log|Z_\text{extra}|^2\big|_{\mu=0}&=\frac{1}{32}(q_{\text{UV}}^2+\bar{q}_{\text{UV}}^2)+O(q_{\text{UV}}^{4},\bar{q}_{\text{UV}}^{4})\\
&=8(q+\bar{q})+O(q^{2},\bar{q}^{2})\,,\\
-\partial_{\mu_1}^4 \log|Z_\text{extra}|^2\big|_{\mu=0} &=O(q_{\text{UV}}^{4},\bar{q}_{\text{UV}}^{4})=O(q^{2},\bar{q}^{2})\,,
\end{split}
\end{align}
where we remind the reader that we are using $q=e^{2\pi i \tau}$, with $\tau$ related to $\tau_{\text{UV}}$ by \eqref{UVtoIR}. The result precisely matches the instanton terms in \eqref{Ffinal}. 

What remains to be justified is the fact that $Z_{\text{inst}}$ gives exponentially suppressed contributions in the large $N$ limit.
Let us start with one instanton,
\begin{align}
Z_{\text{inst}}=1+\frac{\mathtt{a}_1}{2}\,q_{\text{UV}}\prod_{i=1}^4\mu_i + O(q^2_\text{UV})\,,\quad
\mathtt{a}_1=\prod_{l=1}^N \frac{x_l^2}{1+x_l^2}\,,
\end{align}
which only contributes to the derivative of the free energy with all different masses. This is
\begin{align}
-\partial_{\mu_1}\partial_{\mu_2}\partial_{\mu_3}\partial_{\mu_4}F\big|_{\mu=0}=-\frac{1}{2}\langle\langle \mathtt{a}_1\rangle \rangle\,(q_{\text{UV}}+\bar{q}_{\text{UV}})+O(q_{\text{UV}}^3,\bar{q}_{\text{UV}}^3)\,,
\end{align}
where we have defined
\begin{align}
\langle\langle O\rangle\rangle=\frac{\int [dX]e^{-\frac{8\pi^2}{g^2_{\text{YM}}}\text{tr}X^2}e^{-S^{(0)}}\,O}{\int [dX]e^{-\frac{8\pi^2}{g^2_{\text{YM}}}\text{tr}X^2}e^{-S^{(0)}}}=\frac{\langle e^{-S^{(0)}}\,O\rangle}{\langle e^{-S^{(0)}}\rangle}\,.
\end{align}
To compute this, it is convenient to write 
\begin{align}
 \mathtt{a}_1=e^{\mathtt{A}_1}\,,\quad \mathtt{A}_1=\sum_{l=1}^N\left[\log x^2_l-\log (1+x_l^2)\right]\,,
\end{align}
so that at leading order in the large $N$ expansion one has
\begin{align}
\langle\langle \mathtt{a}_1\rangle\rangle\approx e^{\langle\mathtt{A}_1\rangle}\,.
\end{align}
The expectation value of both terms can be computed, at leading order, using Wigner's semi-circle law, with the result that
\es{supp1inst}{
\langle\mathtt{A}_1\rangle\approx \log \frac{\lambda_\text{UV}}{8\pi^2+\lambda_\text{UV}+4\pi \sqrt{\lambda_\text{UV}+4\pi^2}}+\frac{4\pi}{\lambda_\text{UV}}(2\pi-\sqrt{4\pi^2+\lambda_\text{UV}})=-\frac{8\pi}{g_{\text{YM}}}\sqrt{N}+O(1)\,.
}
It will be important that this suppression comes purely from the $\sum_{l=1}^N \log (1+x_l^2)$ part of $\mathtt{A}_1$. In the two instanton calculation, which is also needed to confirm \eqref{Ffinal2}, there is a choice to be made because of the integration over $\phi$. We have checked that performing this integral and then taking the large $N$ limit produces $O(e^{-\sqrt{N}})$ but we believe the mechanism for this can be seen more easily by doing the $\phi$ integration last \cite{Chester:2019jas}. Using similar notation, let
\es{supp2inst}{
\mathtt{a}_2 \equiv \mathtt{a}^{(0)}_2 + \mathtt{a}^{(2)}_2 m_{(2)} + \mathtt{a}^{(4)}_2 (m_{(4)} - 4m_{(2)}^2)\,,
}
where $\mathtt{a}_2$ is the coefficient of $q_\text{UV}^2$ in $Z_\text{inst}$. Looking at the massless contribution and setting $\epsilon_{1,2} = 1 \pm \epsilon$, we have
\es{suppInt}{
\mathtt{a}^{(0)}_2 = \int \frac{d\phi}{2\pi i} \frac{\epsilon^2 \phi^8 / (\epsilon^2 - 1)}{(4\phi^2 - 1)^2[4\phi^2 - (1 + \epsilon)^2][4\phi^2 - (1 - \epsilon)^2]} \frac{\phi^{4N}}{(\phi^2 - 1)^{2N}} e^{\mathtt{A}^{(0)}_2}\,,
}
where
\es{scalePhi}{
\left < \texttt{A}^{(0)}_2 \right > &= \sum_{l = 1}^N \left < 2\log \left ( 1 + \frac{x_l^2}{\phi^2} \right ) - \log \left ( 1 + \frac{x_l^2}{(\phi + 1)^2} \right ) - \log \left ( 1 + \frac{x_l^2}{(\phi - 1)^2} \right ) \right > \\
&= \frac{8\pi}{g_\text{YM}} \sqrt{N} (2|\phi| - |\phi + 1| - |\phi - 1|) + O(1)\,.
}
In other words, there are three terms where each one is a $\phi$-dependent rescaling of what we have computed in \eqref{supp1inst}. The appearance of $-\frac{8\pi}{g_{\text{YM}}}\sqrt{N}$ again now follows from the fact that $\phi = (1 \pm \epsilon)/2$ are the only residues contributing to the integral when $\epsilon \to 0$.
We have shown that, at least up to two instantons, $Z_{\text{inst}}$ contributes to the partition function with terms of order $O(e^{-\sqrt{N}})$, which are non-perturbative contributions in the large $N$ limit and which we can therefore neglect. Based on the simple $N$ dependence of our conjectural results \eqref{Ffinal}, it is tempting to conjecture that this property holds in general. Moreover, this is supported by the derivation of \eqref{scalePhi} since the eigenvalue dependent $\phi_I$ poles of \eqref{z_kinst} take a universal form for all $k$. The function $f(\phi) = 2|\phi| - |\phi + 1| - |\phi - 1|$ is negative for $-1 < \phi < 1$ and for even $k$, every chain of residues for the $\phi_I$ will need to start from one that approaches $1/2$ as $\epsilon_{1,2} \to 1$. For odd $k$, it is possible for a chain to start from a residue of $1$ but in this case the suppression will come from the additional factor of $\sum_{l=1}^N \log (1+x_l^2)$ seen in \eqref{z_kinst}.

In summary, the simplicity of \eqref{Ffinal} has been derived for $k \leq 2$ and there is strong evidence supporting it for all $k$. It is equivalent to the statement that
one can completely neglect $Z_{\text{inst}}$, with the whole instanton contribution to the partition function coming from $Z_{\text{extra}}$.

\section{Flat space amplitudes}
\label{8damp}
In this appendix we derive the flat space scattering amplitudes corresponding to the brane setup that we are interested in. The ten-dimensional string theory setup we are interested in is given by open strings attached to a D7 brane and closed strings propagating in a ten-dimensional bulk. At low energies, this corresponds to 8d gluons and 10d gravitons. We are interested in scattering amplitudes between 8d gluons: note that in this case momentum will be conserved only in eight dimensions, with gluons forced to propagate along the D7 brane with momenta $p_{\parallel}$ while the gravitons can have orthogonal components $p_{\perp}$ to their momenta, since they propagate in the whole ten-dimensional bulk. The action for the model is given by
\begin{align}\label{Stotal}
\begin{split}
S=S_{\text{sugra}}+S_{\text{DBI}}=&\frac{1}{2\kappa^2}\int d^{10}x \sqrt{-g}\left[R+\ldots\right]\\
&-\mu_7\int d^8 y\,\text{tr}\left[g_s^{-1}\sqrt{-\det(\gamma_{\mu\nu}+2\pi\alpha'\,F_{\mu\nu})}+\ldots\right]\,,
\end{split}
\end{align}
where $x$ are 10d coordinates and $y$ are coordinates on the D7 branes, with $g_{MN}$ the 10d bulk metric and $\gamma_{\mu\nu}$ its pull-back to the worldvolume of the D7 branes. The $\ldots$ denote additional terms that are fixed by supersymmetry.  In terms of string theory parameters,  the gravitational coupling constant $\kappa$ is
\begin{align}\label{kappadef}
\kappa^2=8\pi G_N^{(10)}\,,\qquad  G_N^{(10)}=8\pi^6g_s^2\ell_s^8\,,
\end{align}
while the parameter $\mu_7$ appearing in \eqref{DBI_appendix} is the tension of the D7 brane, which in our setup is half of the usual value due to the orientifold, as explained in \cite{Aharony:1999rz}:
\begin{align}
\mu_7=\frac{1}{2(2\pi)^7(\alpha')^4}=\frac{1}{256\pi^6\ell_s^2}\,.
\end{align}

When computing scattering amplitudes, we are going to work in the low energy expansion at small $\alpha'$, where the DBI action reduces (at leading order) to the SYM action. We will then borrow results for SYM amplitudes from the literature, where they are expressed in terms of the 8d YM coupling $\mathtt{g}^2_{\text{8d}}$ appearing in a canonically normalized 8d YM action
\begin{align}\label{YMappendix}
S_{YM}=\frac{1}{4\mathtt{g}^2_{\text{8d}}}\int d^8x \sqrt{-\gamma}F^A_{\mu\nu}F^{A,\mu\nu}\,.
\end{align}
Expanding the Dirac-Born-Infeld (DBI) action in \eqref{Stotal} for small $\alpha'$ one finds
\begin{align}\label{DBI_appendix}
S_{D7}=-\mu_7\int d^8 x \,\text{tr}\left[g_s^{-1}\sqrt{-\det(\gamma_{\mu\nu}+2\pi\alpha'\,F_{\mu\nu})}\right]=\frac{\mu_7}{4g_s}(2\pi\alpha')^2\int d^8 x\sqrt{-\gamma}\, \text{tr}(F_{\mu\nu}F^{\mu\nu})+\ldots\,,
\end{align}
where remember that the string coupling $g_s$ is constant in our setup. Normalizing the gauge group generators with
\begin{align}\label{gennorm_xi}
\text{tr}(T^AT^B)=\xi \,\delta^{AB}\,,
\end{align}
comparing \eqref{YMappendix} and \eqref{DBI_appendix} we then find
\begin{align}\label{gYM_xi}
\mathtt{g}^2_{\text{8d}}=\frac{64\pi^5}{\xi}\,g_s\,\ell_s^4\,,
\end{align}
where we have replaced $\alpha'$ with the string length $\ell_s=\sqrt{\alpha'}$. Note that in our conventions $\xi=2$ so we shall set henceforth
\begin{align}\label{gYM_fixed}
\mathtt{g}^2_{\text{8d}}={32\pi^5}\,g_s\,\ell_s^4\,,
\end{align}

We would would like to compute gluon scattering amplitude in maximally supersymmetric 8d YM theory. Fortunately, this job was already done for us in \cite{Bern:1998ug}, where the answer is given both at tree level and at 1-loop. We note that these results are given for unitary gauge groups $SU(N_c)$, and intermediate steps require relations between partial amplitudes which only apply to unitary groups. However, the final results only depend on the structures $\mathtt{c}_s$, $\mathtt{c}_t$ and $\mathtt{c}_u$ introduced in \cite{Alday:2021odx} (see footnote \eqref{csetc}) at tree level, and on $\mathtt{d}_{st}$, $\mathtt{d}_{su}$ and $\mathtt{d}_{tu}$ introduced in \cite{Alday:2021ajh} (see footnote \ref{dstetc}) at one loop. Hence, we shall borrow the $SU(N_c)$ results and apply them to our case with gauge group $SO(8)$. 

For tree-level amplitudes, we have
\begin{align}
\begin{split}
\mathcal{A}_4^{\text{tree}}=&\mathtt{g}^2_{\text{8d}}\sum_{\sigma \in S_4/\mathbb{Z}_4}\text{tr}(T^{A_{\sigma(1)}}T^{A_{\sigma(2)}}T^{A_{\sigma(3)}}T^{A_{\sigma(4)}})\,A_4^{(0)}(\sigma(1),\sigma(2),\sigma(3),\sigma(4))\\
=&2\mathtt{g}^2_{\text{8d}}\,[\text{Re}\,\text{tr}(T^AT^BT^CT^D)\,A_4^{(0)}(1,2,3,4)+\text{Re}\,\text{tr}(T^AT^CT^DT^B)\,A_4^{(0)}(1,3,4,2)\\
&+\text{Re}\,\text{tr}(T^AT^DT^BT^C)\,A_4^{(0)}(1,4,2,3)]\\
=&\mathtt{g}^2_{\text{8d}}\,[\mathtt{c}_s\,A_4^{(0)}(1,3,4,2)-\mathtt{c}_t\,A_4^{(0)}(1,4,2,3)]\,,
\end{split}
\end{align}
where in the last step we have dropped a term proportional to 
\begin{align}
A_4^{(0)}(1,2,3,4)+A_4^{(0)}(1,3,4,2)+A_4^{(0)}(1,4,2,3)=0\,,
\end{align}
which is known as photon decoupling identity, which can also be verified from the explicit expressions
\begin{align}
\begin{split}
A_4^{(0)}(1,2,3,4)=-\frac{(u+s/w)^2}{s t}\,, & \quad
A_4^{(0)}(1,3,4,2)=-\frac{(u+s/w)^2}{s u}\,,\\
A_4^{(0)}(1,4,2,3)&=-\frac{(u+s/w)^2}{t u}\,,
\end{split}
\end{align}
where we have extracted an overall polarization prefactor of $(\epsilon_1\cdot \epsilon_2)(\epsilon_3\cdot \epsilon_4)$ and introduced the cross-ratio
\begin{align}
w=\frac{(\epsilon_1\cdot \epsilon_2)(\epsilon_3\cdot \epsilon_4)}{(\epsilon_1\cdot \epsilon_3)(\epsilon_2\cdot \epsilon_4)}\,.
\end{align}
The final result reads
\begin{align}
\mathcal{A}_4^{\text{tree}}=(u+s/w)^2\frac{(2\pi)^5g_s\ell_s^4}{s t u}(\mathtt{c}_t\,s-\mathtt{c}_s\,t)\,.
\end{align}
This fixes the overall normalization of the Veneziano amplitude, while higher order terms have the well-known expression given in \eqref{V2}.

At one loop, the $SU(N_c)$ computation from \cite{Bern:1998sv} gives
\begin{align}
\begin{split}
\mathcal{A}_{F^2|F^2}=&2N_c\,\mathtt{g}^4_{\text{8d}}\,[\text{Re}\,\text{tr}(T^AT^BT^CT^D)\,A_4^{(1;1)}(1,2,3,4)+\text{Re}\,\text{tr}(T^AT^CT^DT^B)\,A_4^{(1;1)}(1,3,4,2)\\
&+\text{Re}\,\text{tr}(T^AT^DT^BT^C)\,A_4^{(1;1)}(1,4,2,3)]\\
&+\mathtt{g}^4_{\text{8d}}\,[\text{tr}(T^AT^B)\text{tr}(T^CT^D)\,A_4^{(1;0)}(1,2,3,4)+\text{tr}(T^AT^C)\text{tr}(T^DT^B)\,A_4^{(1;0)}(1,3,4,2)\\
&+\text{tr}(T^AT^D)\text{tr}(T^BT^C)\,A_4^{(1;0)}(1,4,2,3)]\\
=&\mathtt{g}^4_{\text{8d}}\,[\mathtt{d}_{st}\,A_4^{(1;1)}(1,2,3,4)+\mathtt{d}_{su}\,A_4^{(1;1)}(1,3,4,2)+\mathtt{d}_{tu}\,A_4^{(1;1)}(1,4,2,3)]\,,
\end{split}
\end{align}
where we have used the identities  \cite{Bern:1990ux,Naculich:2011ep} 
\begin{align}
\begin{split}
A_4^{(1;0)}(1,2,3,4)&=A_4^{(1;0)}(1,3,4,2)=A_4^{(1;0)}(1,4,2,3)\\
&=2[A_4^{(1;1)}(1,2,3,4)+A_4^{(1;1)}(1,3,4,2)+A_4^{(1;1)}(1,4,2,3)]\,,
\end{split}
\end{align}
as well as the $SU(N_c)$ relations
\begin{align}
\begin{split}
\mathtt{d}_{st}=&2[\text{tr}(T^AT^B)\text{tr}(T^CT^D)+\text{tr}(T^AT^C)\text{tr}(T^DT^B)+\text{tr}(T^AT^D)\text{tr}(T^BT^C)]\\
&+2N_c\,\text{Re}\,\text{tr}(T^AT^BT^CT^D)\,,
\end{split}
\end{align}
and permutations thereof, where the structures $\mathtt{d}_{st}$ etc. are defined as in \cite{Alday:2021ajh}. Finally, from eq. (3.11) of \cite{Bern:1998sv} we read the MHV amplitude at 1-loop
\begin{align}
A^{(1,1)}_{4;4}(1^-,2^-,3^+,4^+)=\frac{i}{4}\frac{\langle 12\rangle^4}{\langle 12\rangle\langle 23\rangle\langle 34\rangle\langle 41\rangle}\text{tr}[p_1p_2p_3p_4]\,\mathcal{I}_4(s,t)\,,
\end{align}
where 
\begin{align}
\text{tr}[p_1p_2p_3p_4]=2^{D/2}\,[(s/2)^2+(t/2)^2-(u/2)^2]=-8\,s\,t\,,
\end{align}
the $2^{D/2}=16$ factor in $D=8$ comes from a trace over gamma matrices, while
\begin{align}
\mathcal{I}_4(s,t)=\int\frac{d^D\ell}{i\,(2\pi)^D}\frac{1}{\ell^2(\ell+p_1)^2(\ell+p_1+p_2)^2(\ell+p_1+p_2+p_3)^2}\,,
\end{align}
is a massless box integral in $D$ dimensions (again we are interested in $D=8$), where all external momenta $p_i$ are taken as incoming. Following, {\it e.g.}, \cite{Valtancoli:2011kr} we can express this as
\begin{align}
\begin{split}
&\mathcal{I}_4(s,t)=\frac{(16\pi)^{(3-D)/2}}{\Gamma[\tfrac{D-3}{2}]}\frac{1}{(D-4)\sin(\tfrac{\pi D}{2})\,s^3\,t^3}\\
&\times\left(s^2(-t)^{D / 2}{ }_2 F_1\left(1, \frac{D}{2}-2 ; \frac{D}{2}-1 ; \frac{s+t}{s}\right)+t^2(-s)^{D / 2}{ }_2 F_1\left(1, \frac{D}{2}-2 ; \frac{D}{2}-1 ; \frac{s+t}{t}\right)\right)\,,
\end{split}
\end{align}
and in the limit $D\to 8$ one has
\begin{align}
\begin{split}
\lim_{D\to 8}\mathcal{I}_4(s,t)&=-\frac{1}{3072\pi^4}\left(f_{\text{box}}^{\text{8d}}(s,t)-2\beta\right)\,,
\end{split}
\end{align}
where $f_{\text{box}}^{\text{8d}}(s,t)$ was introduced in \eqref{fbox8d} and $\beta$ is a (divergent) constant will be regularized by Planck length as we discuss soon. At the end of the day,  using \eqref{gYM_fixed} we find the result quoted in \eqref{Aloop}:
\begin{align}
\mathcal{A}^{\text{1-loop}}_{4}=-\frac{2\pi^6g_s^2\ell_s^8}{3}(u+s/w)^2\left(\mathtt{d}_{st}f_{\text{box}}^{\text{8d}}(s,t)+\mathtt{d}_{su}f_{\text{box}}^{\text{8d}}(s,u)+\mathtt{d}_{tu}f_{\text{box}}^{\text{8d}}(t,u)\right)\,.
\end{align}

Finally, we also have a contribution from the graviton exchange between gluons, which as discussed in \cite{Chester:2023qwo} (adapted to our case) gives
\es{gravitonexch}{
\cA_4^{\text{grav}}&=\,\kappa^2\,(u+s/w)^2\left[\mathtt{t}_1 \int_{\mathbb{R}^2} \frac{d^2p_{\perp}}{(2\pi)^2}\frac{1}{-s+p^2_{\perp}}+\text{permutations}\right]\\
&=-16(u+s/w)^2\,\pi^6g_s^2\ell_s^8\,\left[\mathtt{t}_1(\log(- s)+\ldots)+\text{permutations}\right]\,,
}
where the gravitational coupling constant was introduced in \eqref{kappadef}, the $\ldots$ denote a divergent constant that will be regularized by the Planck length as we discuss next, and note that we integrated over the two transverse dimensions.

We consider the following check of the normalization of the amplitudes given in this section. Note that both the one-loop amplitude \eqref{Aloop} and the graviton exchange amplitude \eqref{gravitonexch} contain logarithms of Mandelstam variables, which are dimensionful quantities. Hence, each such term must appear alongside with $\log \ell_s^2$ terms such as for instance $\log(-s)$ is turned into $\log(-\ell_s^2 s)$ and so on. This is equivalent to the statement that the logarithmic divergences in the above amplitudes are regularized by Planck length.

Now, replacing $\ell_s$ with
\begin{align}
\ell_s=\ell_P\,g_s^{-1/4}\,,
\end{align}
with $\ell_P$ the Planck length, and focusing on the $\log g_s$ dependence, the terms proportional to $\log g_s$ in \eqref{Aloop} and \eqref{gravitonexch} combine into 
\begin{align}
\mathcal{A}_4^{\log}\equiv \left.\mathcal{A}_4^{\text{1-loop}}+\mathcal{A}_4^{\text{grav}}\right|_{\log g_s}=16\pi^6\,\ell_P^8\,(u+s/w)^2\,(\mathtt{t}_1+\mathtt{t}_2+\mathtt{t}_3)\,\log g_s\,.
\end{align}
On the other hand, the small $\ell_s$ expansion of the Veneziano amplitude \eqref{V2} gives rise, at order $\ell_s^2$, to a term which, once $\ell_s$ is converted to $\ell_P$ and $g_s$, reads 
\begin{align}
\mathcal{A}_4^{\text{pert}}\equiv \left. \mathcal{A}_4^{\text{Veneziano}}\right|_{\mathcal{O}(\ell_s^2)}=16\pi^6\,\ell_P^8\,(u+s/w)^2\,(\mathtt{t}_4-2\mathtt{t}_5+3\mathtt{t}_6)\frac{\pi}{3g_s}\,.
\end{align}
Note that the two terms appear with the same power of $\ell_P$ and combining them we read that the dependence on $g_s$ of the sum of the two amplitudes is
\begin{align}
\mathcal{A}_4^{\log}+\mathcal{A}_4^{\text{pert}}=16\pi^6\,\ell_P^8\,(u+s/w)^2\,\bar{\mathcal{A}}(g_s)\,,
\end{align}
where
\begin{align}
\bar{\mathcal{A}}(g_s)=(\mathtt{t}_1+\mathtt{t}_2+\mathtt{t}_3)\,\log g_s+(\mathtt{t}_4-2\mathtt{t}_5+3\mathtt{t}_6)\frac{\pi}{3g_s}\,.
\end{align}
This can be written in terms of projectors and the contributions of the various irreducible representations can be conveniently arranged in the three combinations
\begin{align}\label{barAvcs}
\begin{split}
\bar{\mathcal{A}}_v&=\frac{1}{252}\left[3\bar{\mathcal{A}}_{\mathbf{1}}+60\bar{\mathcal{A}}_{\mathbf{300}}-7\bar{\mathcal{A}}_{\mathbf{35_v}}+14\bar{\mathcal{A}}_{\mathbf{35_c}}+14 \bar{\mathcal{A}}_{\mathbf{35_s}}\right]=-\log\tau_2\,,\\
\bar{\mathcal{A}}_c&=\frac{1}{252}\left[3\bar{\mathcal{A}}_{\mathbf{1}}+60\bar{\mathcal{A}}_{\mathbf{300}}+14\bar{\mathcal{A}}_{\mathbf{35_v}}-7\bar{\mathcal{A}}_{\mathbf{35_c}}+14 \bar{\mathcal{A}}_{\mathbf{35_s}}\right]=-\log\tau_2+\frac{\pi}{2}\tau_2\,,\\
\bar{\mathcal{A}}_s&=\frac{1}{252}\left[3\bar{\mathcal{A}}_{\mathbf{1}}+60\bar{\mathcal{A}}_{\mathbf{300}}+14\bar{\mathcal{A}}_{\mathbf{35_v}}+14\bar{\mathcal{A}}_{\mathbf{35_c}}-7 \bar{\mathcal{A}}_{\mathbf{35_s}}\right]=-\log\tau_2+\frac{\pi}{2}\tau_2\,,
\end{split}
\end{align}
where we have used
\begin{align}
\tau_s=\tau_1+i\,\tau_2=\frac{C_0}{2\pi}+\frac{i}{g_s}\,.
\end{align}
Note that the three combinations \eqref{barAvcs} match the perturbative expansion of quantities that transform into each other under modular transformations, since they have (up to an overall factor) the same $\tau_2$ dependence as \eqref{largeNpert}. Any relative coefficient between the Veneziano and one-loop amplitude would have spoiled this property.

\section{Alternative integration technique}
\label{integration}
Our integrated correlator results \eqref{Is} are based on the Mellin space expression \eqref{Imel}. Here we show how some of the simpler expressions in \eqref{Is} can also be obtained analytically from the position space expression \eqref{d4FCombinedAgain} without numerical integration.

Let us focus on the contact term for illustrative purposes. Inserting
\es{Ddef}{
\bar{D}_{1,1,1,1}(U,V) = \int \frac{ds' dt'}{(4\pi i)^2}U^{s'/2}V^{t'/2} \Gamma[-s'/2]^2\Gamma[-t'/2]^2\Gamma[1 + (s' + t')/2]^2
}
and the inverse Mellin transform of $1$ into the position space integral \eqref{d4FCombinedAgain} yields
\es{doubleMel}{
I[1] &= \frac{1}{64} \int \frac{ds dt ds' dt'}{(2\pi i)^4} \frac{\Gamma \left [ \frac{s + s'}{2} \right ] \Gamma \left [ \frac{t + t'}{2} \right ] \Gamma \left [ \frac{4 - s - t - s' - t'}{2} \right ]}{\Gamma \left [ 2 - \frac{s + s'}{2} \right ] \Gamma \left [ 2 - \frac{t + t'}{2} \right ] \Gamma \left [ \frac{s + t + s' + t' }{2} \right ]} \\
& \times \Gamma \left [ 2 - \frac{s}{2} \right ]^2 \Gamma \left [ 2 - \frac{t}{2} \right ]^2 \Gamma \left [ \frac{s + t - 2}{2} \right ]^2 \Gamma \left [ -\frac{s'}{2} \right ]^2 \Gamma \left [ -\frac{t'}{2} \right ]^2 \Gamma \left [ \frac{s' + t' + 2}{2} \right ]^2
}
where the Feynman integral over $R$ and $\theta$ has been done following \cite{Chester:2020dja}. The gamma functions in the denominator make this difficult but notice that they can be removed by undoing the first Barnes lemma. Using
\es{barnes1}{
\frac{\Gamma(2 - x)^2 \Gamma(-y)^2}{\Gamma(2 - x - y)} = \int \frac{dr}{2\pi i} \Gamma(r)^2 \Gamma(2 - x - r) \Gamma(-y - r)
}
thrice and doubling each Mandelstam variable turns \eqref{doubleMel} into
\es{manyMel}{
I[1] &= \frac{1}{4} \int \frac{dr_1 dr_2 dr_3 ds dt ds' dt'}{(2\pi i)^7} \Gamma(r_1)^2 \Gamma(r_2)^2 \Gamma(r_3)^2 \Gamma(s + s') \Gamma(t + t') \Gamma(2 - s - t - s' - t') \\
& \Gamma(2 - s - r_1) \Gamma(-s' - r_1) \Gamma(2 - t - r_2) \Gamma(-t' - r_2) \Gamma(s + t - 1 - r_3) \Gamma(1 + s' + t' - r_3).
}
We can now evaluate the $s, t, s', t'$ integrals with the first Barnes lemma and apply the duplication formula to find
\es{dup1}{
I[1] &= \frac{1}{4} \int \prod_{i = 1}^3 \frac{dr_i}{2\pi i} \Gamma(r_i)^2 \Gamma(2 - 2r_i) \frac{\Gamma(1 - r_1 - r_2 - r_3) \Gamma(3 - r_1 - r_2 - r_3)}{\Gamma(6 - 2r_1 - 2r_2 - 2r_3)} \\
&= \frac{1}{128} \int \prod_{i = 1}^3 \frac{dr_i}{2\pi i} 4^{r_i} \Gamma(r_i)^2 \Gamma(2 - 2r_i) \frac{\sqrt{\pi} \Gamma(1 - r_1 - r_2 - r_3)}{\Gamma(\tfrac{7}{2} - r_1 - r_2 - r_3)}.
}
After inserting the integral representation of the beta function, we can again use the duplication formula on the factored $r_i$ integrals.
\es{dup2}{
I[1] &= \frac{1}{96} \int_0^1 dz (1 - z)^{\frac{3}{2}} \int \prod_{i = 1}^3 \frac{dr_i}{2\pi i} \left ( \frac{4}{z} \right )^{r_i} \Gamma(r_i)^2 \Gamma(2 - 2r_i) \\
&= \frac{1}{12} \pi^{-\frac{3}{2}} \int_0^1 dz (1 - z)^{\frac{3}{2}} \int \prod_{r = 1}^3 \frac{dr_i}{2\pi i} \Gamma(r_i)^2 \Gamma(1 - r_i) \Gamma(\tfrac{3}{2} - r_i) z^{-r_i}
}
The last step is to use the integral representation of a hypergeometric function which thankfully leads to the closed-form expression
\es{analyticContact}{
I[1] = \frac{1}{324} \int_0^1 dz (1 - z)^{\frac{3}{2}} {}_2F_1(1; \tfrac{3}{2}; \tfrac{5}{2}; 1 - z)^3 = \frac{1}{24}\,.
}

\bibliographystyle{JHEP}
\bibliography{Ftheory}

\end{document}